\documentclass[aip,reprint]{revtex4-1}
\usepackage{graphicx,amsmath,amssymb,color,amsthm}
\usepackage{bigints}
\usepackage{relsize}
\newtheorem{definition}{Definition}

\begin{document}


\title{Quantifying chaos using Lagrangian descriptors} 



\author{M.~Hillebrand}
\email[]{m.hillebrand@uct.ac.za}
\affiliation{Nonlinear Dynamics and Chaos Group, Department of Mathematics and
Applied Mathematics, University of Cape Town, Rondebosch 7701,
South Africa.}

\author{S.~Zimper}
\affiliation{Nonlinear Dynamics and Chaos Group, Department of Mathematics and
Applied Mathematics, University of Cape Town, Rondebosch 7701,
South Africa.}

\author{A.~Ngapasare}
\affiliation{Nonlinear Dynamics and Chaos Group, Department of Mathematics and
Applied Mathematics, University of Cape Town, Rondebosch 7701,
South Africa.}

\author{M.~Katsanikas}
\affiliation{Research Center for Astronomy and Applied Mathematics, Academy of Athens, Soranou Efesiou 4, Athens, GR-11527, Greece.}

\author{S.~Wiggins}
\affiliation{School of Mathematics, University of Bristol, Fry Building, Woodland Road, Bristol, BS8 1UG, United Kingdom.}
\affiliation{Department of Mathematics, 
United States Naval Academy, 
Chauvenet Hall, 572C Holloway Road, 
Annapolis, MD 21402-5002. 
}

\author{Ch.~Skokos}
\affiliation{Nonlinear Dynamics and Chaos Group, Department of Mathematics and
Applied Mathematics, University of Cape Town, Rondebosch 7701,
South Africa.}


\date{\today}

\begin{abstract}
We present and validate simple and efficient methods to estimate the chaoticity of orbits in low-dimensional conservative dynamical systems, namely autonomous Hamiltonian systems and area-preserving symplectic maps, from computations of Lagrangian descriptors (LDs) on short time scales. Two  quantities are proposed for determining the chaotic or regular nature of orbits in a system's phase space, which are based on the values of the LDs of these orbits and of nearby ones: The difference (DNLD) and ratio (RNLD) of neighboring orbits' LDs.
Using as generic test models the prototypical two degree of freedom H{\'e}non--Heiles system and the two-dimensional standard map, we find that these indicators are able to correctly characterize the chaotic or regular nature of orbits to better than 90\% agreement with results obtained by implementing the Smaller Alignment Index (SALI) method, which is a well established chaos detection technique.
Further investigating the performance of the two introduced quantities we discuss the effects of the total integration time and of the spacing between the used neighboring orbits on the accuracy of the methods, finding that even typical short time, coarse grid LD computations are sufficient to provide a reliable quantification of the systems' chaotic component, using less CPU time than the SALI.
In addition to quantifying chaos, the introduced indicators have the ability to reveal details about the systems' local and global chaotic phase space structure.
Our findings clearly suggest that LDs can also be used to quantify and investigate chaos in continuous and discrete low-dimensional conservative dynamical systems.

\end{abstract}


\maketitle
\begin{quotation}
Identifying the chaotic or regular nature of orbits of dynamical systems is a fundamental task of nonlinear dynamics theory. Chaos identification can be done by implementing techniques based on the evolution of the orbits themselves and of small perturbations to these orbits, like the computation of the Smaller Alignment Index (SALI), which is a well established and  efficient chaos indicator. Here we show that the recently introduced method of Lagrangian descriptors (LDs), which allows for the efficient revelation of phase space structures, can also be used to globally investigate the chaoticity of low-dimensional conservative systems. More specifically, we introduce two quantities derived directly from the LD computations of neighboring orbits as coarse-grid indicators capable of distinguishing between chaotic and regular trajectories. We show that these two quantities perform very well when benchmarked against accurate estimations of chaotic behavior from the SALI method, as they  correctly characterize  the chaotic nature  of over $90\%$ of the orbits, despite being based on short time computations. Furthermore, the introduced quantities probe the phase space dynamics, enabling effective visualization of the space's structural properties.
\end{quotation}


\section{Introduction}
\label{sec:intro}

The problem of efficiently characterizing orbits  of dynamical systems as chaotic or regular has been around for over a century. The pioneering work of Lyapunov~\cite{lyapunov1992} who introduced some asymptotic measures to characterize the growth or shrinking of small perturbations  to orbits (usually called `deviation vectors') in dynamical systems cannot be over-emphasized. These measures have over the years been termed Lyapunov exponents (LEs). Oseledets~\cite{oseledec1968} applied LEs to chaotic motion, developing the so-called `multiplicative rrgodic theorem', and provided the theoretical basis for their numerical computation, which was implemented a few years later~\cite{benettin1980,benettin1980b} (see also Ref.~[\onlinecite{skokos2010}] and references therein). The computation of the maximum LE (mLE) is the most commonly used chaos indicator, as a positive value of the index denotes chaotic behavior, and has been applied in studies of such diverse systems as DNA molecules,~\cite{hillebrand2019} graphene nanoribbons,~\cite{hillebrand2020} disordered granular chains~\cite{achilleos2018,ngapasare2019} and soft architected structures,~\cite{ngapasare2022} neural networks,~\cite{wang2020back} as well as for models describing the motion of planetary satellites,~\cite{shevchenko2002} and particle motion in the vicinity of black holes.~\cite{lei2021}

Over the years several chaos indicators have been developed to overcome the problem of the slow convergence of the mLE estimator to its limiting value, such as the Fast Lyapunov Indicator (FLI),~\cite{froeschle1997} the Mean Exponential Growth of Nearby Orbits (MEGNO),~\cite{CS00} the Smaller Alignment Index (SALI),~\cite{skokos2001} and its extension, the so-called Generalized Alignment Index (GALI).~\cite{skokos2007} Review papers on many modern chaos detection techniques can be found in Ref.~[\onlinecite{SGL_LNP_16}], while a detailed comparison of different chaos indicators which utilize deviation vectors is performed in Ref.~[\onlinecite{maffione2011}]. In our study we will use the SALI method in order to accurately reveal the chaotic or regular nature of orbits.
The SALI has proved to be an efficient chaos indicator and has been successfully implemented in studies of various dynamical systems (see for example Ref.~[\onlinecite{SM16}] and references therein). 

From a dynamical systems perspective, it is fundamentally important to characterize a system's chaotic behavior both locally and globally. In the local case, the characterization is made within the vicinity of a particular orbit, while in the global case it is performed over a large set of initial conditions (ICs). An efficient way to visualize the global dynamics of a low-dimensional dynamical system, and in particular a 2 degree of freedom (2dof) autonomous Hamiltonian system, is to construct the related Poincar{\'e} surface of section (PSS), which in that case has 2 dimensions and can be easily visualized (see for example Chapt.~1 in Ref.~[\onlinecite{LL92}]). Scattered points on a PSS indicate chaos, while points forming smooth looking curves belong to regular orbits, similar to what is observed for orbits in the phase space portraits of 2-dimensional (2D) symplectic maps, which are discrete time dynamical systems. In our study we will investigate the chaotic behavior of two prototypical systems, namely the 2dof H{\'e}non--Heiles Hamiltonian~\cite{henon1964} and the 2D standard map,~\cite{chirikov1979} by  respectively combining PSS and phase portrait plots with SALI computations in order to reveal the systems' global dynamics.
H{\'e}non--Heiles

A recently introduced  technique with origins from oceanographic studies,~\cite{madrid2009,mancho2013} the Lagrangian descriptors (LDs) method,~\cite{ldbook2020} has proved useful in revealing and visualizing phase space structures with applications ranging from describing the dynamics of chemical reactions,~\cite{junginger2016,craven2017,Agaoglou2019} unveiling the behavior of molecular structures,~\cite{craven2016,revuelta2019,agaoglou2021} constructing time dependent dividing surfaces,~\cite{feldmaier2017} detecting dynamical matching in a Caldera potential,~\cite{katsanikas2020} investigating the properties of open and unbounded maps,~\cite{carlo2020, garcia2020} finding bifurcations of periodic orbits,~\cite{katsanikas2022bifurcation} 3D vector fields,~\cite{garcia2018} stochastic dynamical systems,~\cite{balibrea2016} as well as of dissipative systems,~\cite{garcia2022} amongst a plethora of other applications. LDs have recently also been used to extract Lagrangian coherent structures in cardiovascular flows.~\cite{darwish2021}
One of the main reasons for the popularity of this method within the dynamical systems community is its simplicity. Recently, a few studies have been reported in the literature which have attempted to make a connection between LDs and chaoticity.~\cite{demian2017,montes2021}

In this paper, we introduce two methods which are based on LD computations to globally characterize the chaotic nature of the dynamics of low-dimensional systems, in a similar way to that which would be revealed by the PSS and the SALI methods. To the best of our knowledge, this is the first time LDs have been directly used as a chaos detection technique. In particular, we demonstrate how quantities based on LDs can be used as chaos indicators within a reasonable degree of accuracy, using as test cases the aforementioned H{\'e}non--Heiles model  and 2D standard map. Our approach has the  advantage of being computationally cheaper than standard methods (like the SALI) based on the evolution of deviation vectors, which require the integration (iteration) of both the equations of motion (the map), as well as of the so-called variational equations (tangent map) governing the evolution of the deviation vector of a Hamiltonian system (symplectic map).~\cite{skokos2010,SM16}

The rest of the paper is organized as follows. In Sect.~\ref{sec:theory} we describe our approach in detail and define the two indicators we use in our work. In Sect.~\ref{sec:numerical} we discuss the numerical implementation of these methods to the H{\'e}non--Heiles system  and the 2D standard map. Finally, we present our conclusions and discuss the significance of our findings in Sect.~\ref{sec:conclusions}.

\section{Numerical techniques} 
\label{sec:theory}

A basic characteristic of chaotic behavior is the sensitive dependence on ICs,~\cite{devaney} which leads to the exponential separation of initially nearby orbits, or in other words to the exponential growth of the length of deviation vectors. Exploiting this feature, several chaos indicators like the  mLE,~\cite{oseledec1968} the FLI,~\cite{froeschle1997} and the MEGNO,~\cite{CS00,cincotta2016} discriminate between regular and chaotic orbits, while the SALI and  GALI techniques make use of the convergence of arbitrary deviation vectors to the direction defined by the mLE.~\cite{skokos2007,SM16}

According to Refs.~[\onlinecite{skokos2001,skokos2004,SM16}] in order to evaluate the SALI of an orbit we follow the evolution of the orbit itself and of two, initially linearly independent, deviation vectors  $\mathbf{v_1}(0)$ and $\mathbf{v_2}(0)$ from it. Then at time $t$ we compute SALI$(t)$ as 
\begin{equation} 
\label{eq:SALI} 
\mbox{SALI}(t)=\min \left\{\|
    \hat{\mathbf{v}}_1(t) + \hat{\mathbf{v}}_2(t) \|, \| \hat{\mathbf{v}}_1(t)
    - \hat{\mathbf{v}}_2(t) \| \right\},
\end{equation}
with $\| \cdot \|$ denoting the usual Euclidean norm and  $\hat{\mathbf{v}}_i(t) = \frac{\mathbf{v}_i(t)}{\| \mathbf{v}_i(t) \|}$, $i=1,2$ being unit vectors. In the case of chaotic orbits, the two deviation vectors will eventually be aligned to the direction defined by the mLE and consequently the SALI will become zero, following an exponential decay which depends on the values of the two largest Lyapunov exponents $\lambda_1 \geq \lambda_2$, i.e.,
\begin{equation}
\label{eq:SALI_ch}\mbox{SALI}(t) \propto e^{-(\lambda_1-\lambda_2)t}.
\end{equation}
On the other hand, in the case of regular orbits, the two deviation vectors will fall on the tangent space of the torus on which the motion takes place. 

In the case of a $2$dof Hamiltonian system, like the  H{\'e}non--Heiles model we consider here, we have $\lambda_2=0$ (see for example Ref.~[\onlinecite{skokos2010}]) and the behavior of SALI is~\cite{SM16} 
\begin{equation}
\label{eq:SALI_HH}
\mbox{SALI}(t) \propto
\begin{cases}
\mbox{constant} & \mbox{for regular orbits} \\
e^{-\lambda_1 t} & \mbox{for chaotic orbits}
\end{cases}. 
\end{equation}
In the case of $2$D symplectic maps, regular motion occurs on 1D curves, whose tangent space is also $1$D. Thus, also in the case of regular motion, the two deviation vectors will eventually have the same direction. Consequently the SALI will vanish, but this will be done with the index following a power law decay, in contrast to the exponential decay observed for chaotic orbits for which $\lambda_2=-\lambda_1$.~\cite{skokos2010} So in this case we get~\cite{SM16}
\begin{equation}
\label{eq:SALI_2Dmap}
\mbox{SALI}(T) \propto
\begin{cases}
\frac{1}{T} & \mbox{for regular orbits} \\
e^{-2\lambda_1 T} & \mbox{for chaotic orbits}
\end{cases}, 
\end{equation}
with $T$ denoting the number of map iterations. 

Here we will utilize  the exponential growth of the phase space distance between initially nearby chaotic orbits to introduce a simple method of estimating chaoticity in low-dimensional systems using LDs. First, we consider an $N$D continuous dynamical system, whose state at time $t$ is defined by an $N$D point $\mathbf{x}(t)$ in the system's phase space and its evolution by a vector field of the form
\begin{equation}
    \label{eq:dynsys}
    \dot{\mathbf{x}}(t) = \frac{d\mathbf{x}(t)}{dt}= \mathbf{f}\left(\mathbf{x}(t;\mathbf{x}_0),t\right),
\end{equation}
with $\mathbf{x}_0 \equiv \mathbf{x}(0)$ being the IC of an orbit,  and $\mathbf{f}$ an $N$D vector function which is differentiable in the phase space coordinates $\mathbf{x}$ and continuous in time. Then, the `$p$-norm' LD  for  $\mathbf{x}_0$ is defined as~\cite{lopesino2017}
\begin{equation}
    \label{eq:defLD}
    M_p(\mathbf{x}_0) = \displaystyle \bigintssss_{-\tau}^{\tau}\sum_{i=1}^{N}\left|f_i\left(\mathbf{x}(t; \mathbf{x}_0),t\right)\right|^p \,\,dt, \qquad 0<p\leq 1,
\end{equation}
with $\tau>0$ determining the length of the considered time window.
Although the quantity defined in \eqref{eq:defLD} is not a norm of the velocity vector $\dot{\mathbf{x}}(t)$, we use the term `$p$-norm’ LD, which was also adopted in previous publications,~\cite{lopesino2017,katsanikas2020,ldbook2020} in order to emphasize the appearance of the $p$ power of the norm of each velocity component $f_i$.
The expression in~\eqref{eq:defLD} can be decomposed into the forward and backward time contributions of the integral, $M_p^{f}$ and $M_p^{b}$ respectively. A similar but subtly different formulation of the LD uses the arc length, simply giving
\begin{align}
    \label{eq:arclength}
    M(\mathbf{x}_0) &= \bigintssss_{-\tau}^{\tau}   \left|\left|\dot{\mathbf{x}}(t)\right|\right| \,\,dt = \bigintssss_{-\tau}^{\tau} \left|\left|\mathbf{f} \left(\mathbf{x}(t; \mathbf{x}_0),t\right)\right|\right|\,\,dt \nonumber\\
    &=\bigintss_{-\tau}^{\tau}\sqrt{\sum_{i=1}^{N} f_i^2 \left(\mathbf{x}(t; \mathbf{x}_0),t\right)} \,\,dt.
\end{align}

The LDs have also been used in discrete time systems. So, for a general $N$D map of the form
\begin{equation}
    \label{eq:dismap}
    \mathbf{x}_{j+1} = \mathbf{g}(\mathbf{x}_j),
\end{equation}
with $\mathbf{x}_{j}$ being an $N$D vector denoting the state of the system at iteration $j$, and $\mathbf{g}$ a general $N$D vector function of the phase space coordinates, a `$p$-norm' formulation,  similar to the one defined for the continuous time case in~\eqref{eq:defLD}, exists~\cite{lopesino2015}
\begin{equation}
    \label{eq:defDisLD}
    MD_p = \mathlarger{\sum}_{j=-T}^{T-1} \mathlarger{\sum}_{i=1}^{N}\left|x^{(i)}_{j+1}-x^{(i)}_{j}\right|^p, \qquad 0< p \leq 1,
\end{equation}
where  $i$ indexes the $N$ elements of the vector $\mathbf{x}$, and $T$ is the number of iterations. Furthermore, the adaptation of the arc length LD \eqref{eq:arclength} to the case of discrete time maps leads to the quantity 
\begin{equation}
    \label{eq:defDisLDarclength}
    MD = \mathlarger{\mathlarger{\sum}}_{j=-T}^{T-1} \sqrt{\mathlarger{\sum}_{i=1}^{N}\left(x^{(i)}_{j+1}-x^{(i)}_{j}\right)^{2}}
\end{equation}

It is worth noting that, in the analysis below, we use the notation of the LD corresponding to the definition~\eqref{eq:arclength} related to the arc length, mainly because this definition is intuitively clearer, but our approach and arguments can similarly be implemented when the `$p$-norm' version of the LD is used. Furthermore, for simplicity reasons, we will refer only to the  case of continuous time systems, but analogous arguments hold for maps. We also emphasize that for all our numerical results we used the `$p$-norm' LD of Eqs.~\eqref{eq:defLD} and \eqref{eq:defDisLD} with $p=0.5$, as this version of the LDs has been recommended for effective detection of dynamical features,~\cite{mancho2013,lopesino2017} and successfully implemented in various studies,~\cite{demian2017, garcia2020, katsanikas2020} and considered only the forward in time contributions to the LDs.

\subsection{The difference of LDs of neighboring orbits} 
\label{sec:DNLP}

Aiming to identify the chaotic or regular nature of an orbit starting at point $\mathbf{x}$ in the phase space of a continuous time dynamical system, let us consider the  forward LD values, $M^f$, of this orbit and of a nearby one, initially located at $\mathbf{x}^\prime = \mathbf{x}+\mathbf{w}$ with $\mathbf{w}$ being a small perturbation. Then, according to \eqref{eq:arclength}, at any time $\tau>0$, the absolute difference of these forward LDs is given by 
\begin{align}
    \label{eq:diff1}
    \displaystyle \left|M^f(\mathbf{x}) - M^f(\mathbf{x}^\prime)\right| &= \displaystyle \left|\int_0^\tau \left(||\dot{\mathbf{x}}||-||\dot{\mathbf{x}}^\prime||\right) \,\,dt \right| \nonumber \\
        &= \displaystyle \left|\int_0^\tau \left( ||\dot{\mathbf{x}}||-||\dot{\mathbf{x}}+\dot{\mathbf{w}}||\right) \,\,dt \right|.
\end{align}
Inherently, this quantity encodes information about the evolution of the deviation between the two orbits, and consequently about the chaotic (exponential growth of the deviation) or regular (polynomial or typically linear increase of the deviation's size~\cite{skokos2007}) nature of the orbit.
Consequently it is reasonable to expect that the difference of the two LDs~\eqref{eq:diff1} will be noticeably larger for chaotic orbits.
Actually, it is more appropriate to consider the difference in arc lengths between these two orbits ($\mathbf{x}$ and $\mathbf{x}^\prime$) in the context of the actual arc length of the reference orbit $||\dot{\mathbf{x}}||$ by considering the ratio   
\begin{align}
    \label{eq:diff2}
    \displaystyle \delta^f (\mathbf{x}) &= \displaystyle \frac{\left|M^f(\mathbf{x}) - M^f(\mathbf{x}^\prime)\right|}{|M^f(\mathbf{x})|} \nonumber \\
        &= \displaystyle \frac{\displaystyle \left|\int_0^\tau \left( ||\dot{\mathbf{x}}||-||\dot{\mathbf{x}}+\dot{\mathbf{w}}||\right) ]\,\,dt \right|}{\displaystyle \int_0^\tau ||\dot{\mathbf{x}}||\,\,dt}.
\end{align}

It is clear that the exponential (or not) growth rate of $\mathbf{w}$ decisively determines the magnitude of $\delta^f(\mathbf{x})$ as $\tau$ grows. In particular, the value of $\delta^f(\mathbf{x})$ for chaotic orbits will eventually be substantially larger than the ones obtained for regular ones, allowing the discrimination between the two cases. The same behavior will be observed if we substitute in Eqs.~\eqref{eq:diff1} and \eqref{eq:diff2} $M^f$ with the backward time LD $M^b$, defining in this way an analogous quantity to $\delta^f (\mathbf{x})$, which we will naturally denote as  $\delta^b (\mathbf{x})$. The important observation here  is that the essential information about the  chaoticity of an orbit is actually equivalently encoded in each one of the forward or backward computation of the LD. Thus,  one of these  LDs is sufficient to reveal the potential chaoticity of an orbit, and consequently be used as a chaos indicator. In particular, in our study we  consider only computations of the forward LDs. Generalizing these ideas to dynamical systems with multidimensional phase spaces, we can infer that quantities similar to the one in \eqref{eq:diff2}, based on LDs computations from several neighboring orbits around the tested orbit $\mathbf{x}$, in many (if not all) possible phase space directions, can be used as effective chaos indicators. Thus, we are led to introduce the following quantity.

\begin{definition} {\bf The Difference of Neighboring orbits' Lagrangian Descriptors (DNLD), \boldmath$D_L^n$.} We consider ICs of orbits on a finite grid of an $n(\geq 1)$-dimensional subspace of the $N(\geq n)$-dimensional phase space of a dynamical system, and the LDs of these orbits. Then, any non-boundary point $\mathbf{x}$ in this subspace has $2n$ nearest neighbors
\begin{equation*}
     \mathbf{y}_i^{\pm} = \mathbf{x}\pm \sigma^{(i)} \mathbf{e}^{(i)}, \,\, i=1,2,\ldots n,
 \end{equation*} 
where $\mathbf{e}^{(i)}$ is the $i^{\textit{th}}$ usual basis vector in $\mathbb{R}^n$, and $\sigma^{(i)}$ is the distance between successive grid points in this direction. If we denote by $M(\mathbf{x})$ and $M\left(\mathbf{y}_i^\pm\right)$ the LDs of these orbits, then the difference of neighboring orbits' Lagrangian descriptors (DNLD) at $\mathbf{x}$ in this subspace is defined as 
\begin{align}
     \label{eq:defDNLD}
     \displaystyle D_L^n(\mathbf{x}) = \frac{1}{2n} \mathlarger{\mathlarger{\sum}}_{i=1}^{n} \frac{\left|M(\mathbf{x}) - M(\mathbf{y}_i^+)\right| +\left|M(\mathbf{x}) - M(\mathbf{y}_i^-)\right|}{M(\mathbf{x})}.
\end{align}
\end{definition}
  
It is apparent that by implementing the DNLD indicator $D_L^n(\mathbf{x})$ \eqref{eq:defDNLD} we are actually evaluating the average of multiple normalized absolute differences of LDs of neighboring orbits. Thus, by combining all these differences we are extracting  information  about the magnitude of growth of `small deviations' in many directions from the studied orbit, getting in this way a more global understanding of the dynamical properties of the orbit's neighborhood.
Due to the exponential growth of perturbations of chaotic orbits,  we expect the  differences between LD values of neighboring orbits to be larger for such orbits than the differences of LDs encountered for regular orbits.
Thus, using the $D_L^n(\mathbf{x})$ index we can classify an orbit with IC   $\mathbf{x}$ as chaotic if $D_L^n(\mathbf{x})\geq \alpha_D$, for some appropriately chosen positive threshold value $\alpha_D$.

\subsection{The ratio of LDs of neighboring orbits} 
\label{sec:RNLP}

Another approach to relate the growth rate of separations between initially nearby orbits to their LDs, in order to construct a quantity which can be used as a chaos indicator, is to consider the ratio of the these LD values,
\begin{equation}
\label{eq:ratio_M}
    \rho^f(\mathbf{x}) = \frac{M^f(\mathbf{x}^\prime)}{M^f(\mathbf{x})} = \frac{\displaystyle \int_0^\tau ||\dot{\mathbf{x}}+\dot{\mathbf{w}}||\,\,dt}{\displaystyle \int_0^\tau ||\dot{\mathbf{x}}||\,\,dt}.
\end{equation}
It is evident that the exponential growth in the magnitude of $\mathbf{w}$, which is encountered in the case of chaotic orbits, will result in a divergence of the $\rho^f(\mathbf{x})$ value away from 1. On the other hand, in the case of regular orbits for which the length of $\mathbf{w}$ increases much more slowly, the growth of the numerator in \eqref{eq:ratio_M} is not expected to be very different to the growth of the denominator, resulting in $\rho^f(\mathbf{x})$ values closer to 1.
Thus, the information about the chaotic or regular nature of an orbit is related to the deviation or the proximity of $\rho^f(\mathbf{x})$ to 1. Although in \eqref{eq:ratio_M} we considered the forward time LD $M^f$, a similar expression can be written for the backward time LD $M^b$, which will exhibit the same behavior as $M^f$ for chaotic and regular orbits. So, we can combine together both the $\rho^f(\mathbf{x})$ and the analogous $\rho^b(\mathbf{x})$ quantities in order to devise an index whose deviation from 1 can be used to identify chaos. Nevertheless, in our numerical investigations we consider results obtained only by the use of the forward LDs, as they require less computational effort and practically lead to the same outcomes.

\begin{definition} {\bf The Ratio of Neighboring orbits' Lagrangian Descriptors (RNLD), \boldmath$R_L^n$.}
Under the same  conditions used for the definition of the  DNLD index $D_L^n$, we introduce the ratio of neighboring orbits' Lagrangian descriptors (RNLD) to be
\begin{equation}
     \label{eq:defRNLD}
     R_L^n(\mathbf{x}) = \left|1-\frac{1}{2n}\mathlarger{\mathlarger{\sum}}_{i=1}^{n}\frac{M(\mathbf{y}_i^+) +M(\mathbf{y}_i^-)}{{M(\mathbf{x})}}\right|.
\end{equation}
\end{definition}

Similarly to the DNLD \eqref{eq:defDNLD} the $R_L^n(\mathbf{x})$ index is also based on information from several nearby orbits to the tested one, capturing in this way the general dynamical properties of the orbit's phase space neighborhood, and can efficiently be used to identify chaos. In particular, an orbit with IC $\mathbf{x}$ is characterized as chaotic if $R_L^n(\mathbf{x})\geq \alpha_R$, for some appropriately chosen threshold value $\alpha_R>0$, while  $R_L^n(\mathbf{x})< \alpha_R$ classifies the orbit as regular.

\section{Numerical results} 
\label{sec:numerical}

In this section we  investigate the ability of both the $D_L^n$  \eqref{eq:defDNLD} and the $R_L^n$ \eqref{eq:defRNLD} indices to distinguish between chaotic and regular motion by applying them to two prototypical, well-known low-dimensional conservative systems of continuous and discrete time, namely the $2$dof H\'enon--Heiles Hamiltonian model \cite{henon1964} and the $2$D standard map.~\cite{chirikov1979}
We note that the rich dynamics evident in these models render them good test systems for nonlinear dynamical techniques, as has been seen in the past.\cite{froeschle1997,skokos2001,skokos2004}
In our study we  present results obtained for individual orbits as well as for ensembles  of orbits in phase space subspaces of different dimensions, based only on the evaluations of the forward LDs.  

\subsection{Detecting chaos in the H{\'e}non--HeilesHeiles system} 
\label{sec:numerical_HH}

The H\'enon--Heiles Hamiltonian is a low-dimensional system whose chaotic behavior has been extensively studied since its introduction in Ref.~[\onlinecite{henon1964}], as a model of the motion of a star at the central regions of the symmetry plane of a galaxy. Its Hamiltonian function is given by
\begin{equation}
    \label{eq:defHH}
    H(x,y,p_x,p_y) = \frac{1}{2} \left(p_x^2 +p_y^2 + x^2 + y^2  \right) + x^2 y - \frac{y^3}{3} , 
\end{equation}
with $x$, $y$ being the coordinates of the star and $p_x$, $p_y$ the conjugate momenta. The system has a 4D phase space but for a fixed value $H$ of the Hamiltonian function (which typically is referred as the system's energy)  its dynamics can be efficiently visualized on the 2D PSS defined by $x=0$ and $p_x>0$. 

\begin{figure}
    \centering
    \includegraphics[width=\columnwidth]{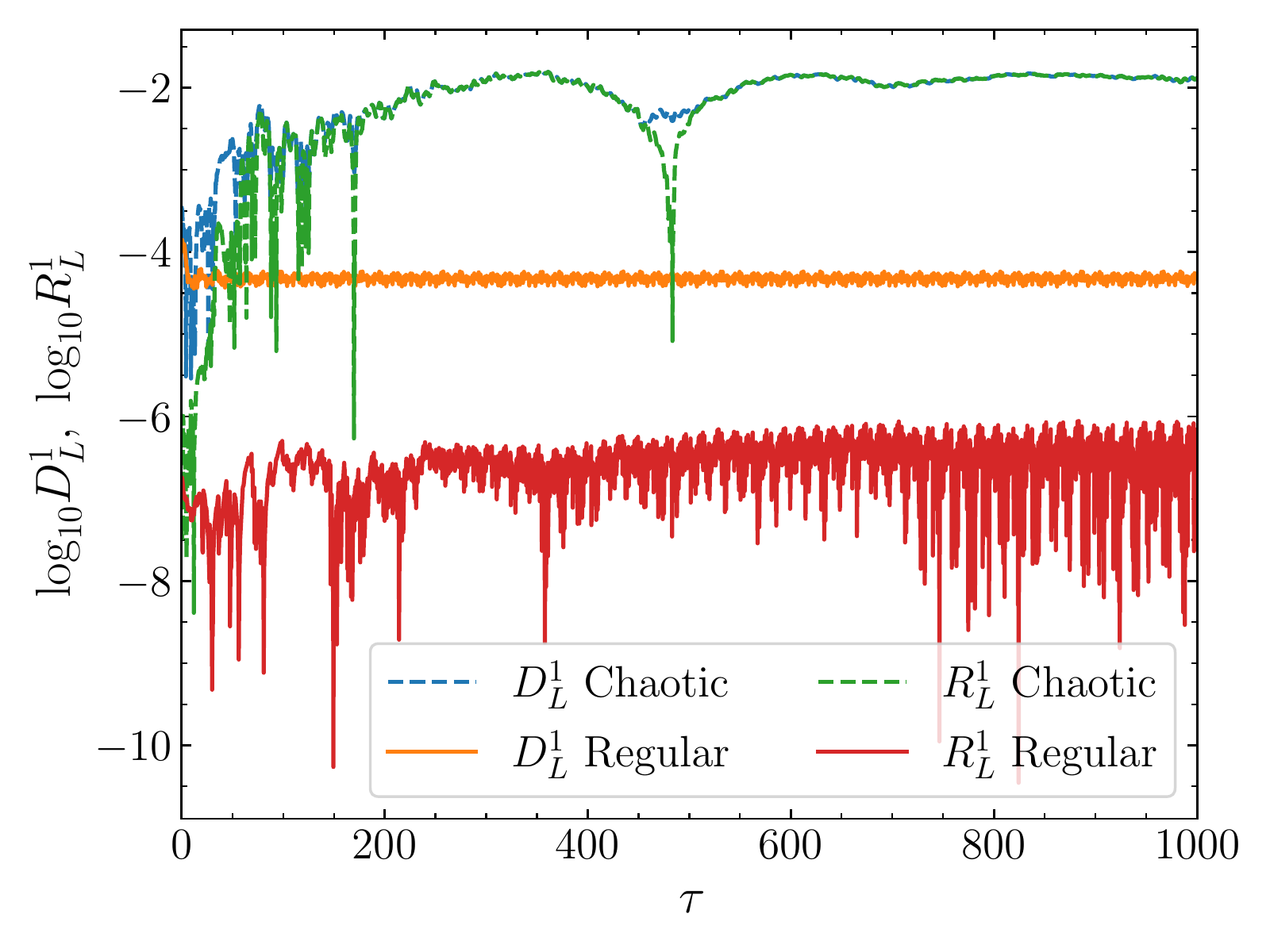}
    \caption{The time evolution of the $D_L^1$  \eqref{eq:defDNLD} and the $R_L^1$ \eqref{eq:defRNLD} indices, based on the computation of  forward LDs in the time interval $[0, \tau]$, for two representative orbits of the H\'enon--Heiles system \eqref{eq:defHH} with $H=1/8$, a chaotic one with IC $x=p_y=0$, $y=-0.15$, $p_x>0$ [$D_L^1$: dashed blue curve, $R_L^1$: dashed green curve], and a regular one with IC $x=p_y=0$, $y=0.2$, $p_x>0$ [$D_L^1$: orange curve, $R_L^1$: red curve]. The ICs of these orbits are denoted by blue (chaotic) and green (regular) points  Fig.~\ref{fig:1d_125_cha_reg}(a). 
    }
    \label{fig:NLD_illustrate}
\end{figure}

\begin{figure*} 
    \centering
    \includegraphics[width=0.9\columnwidth]{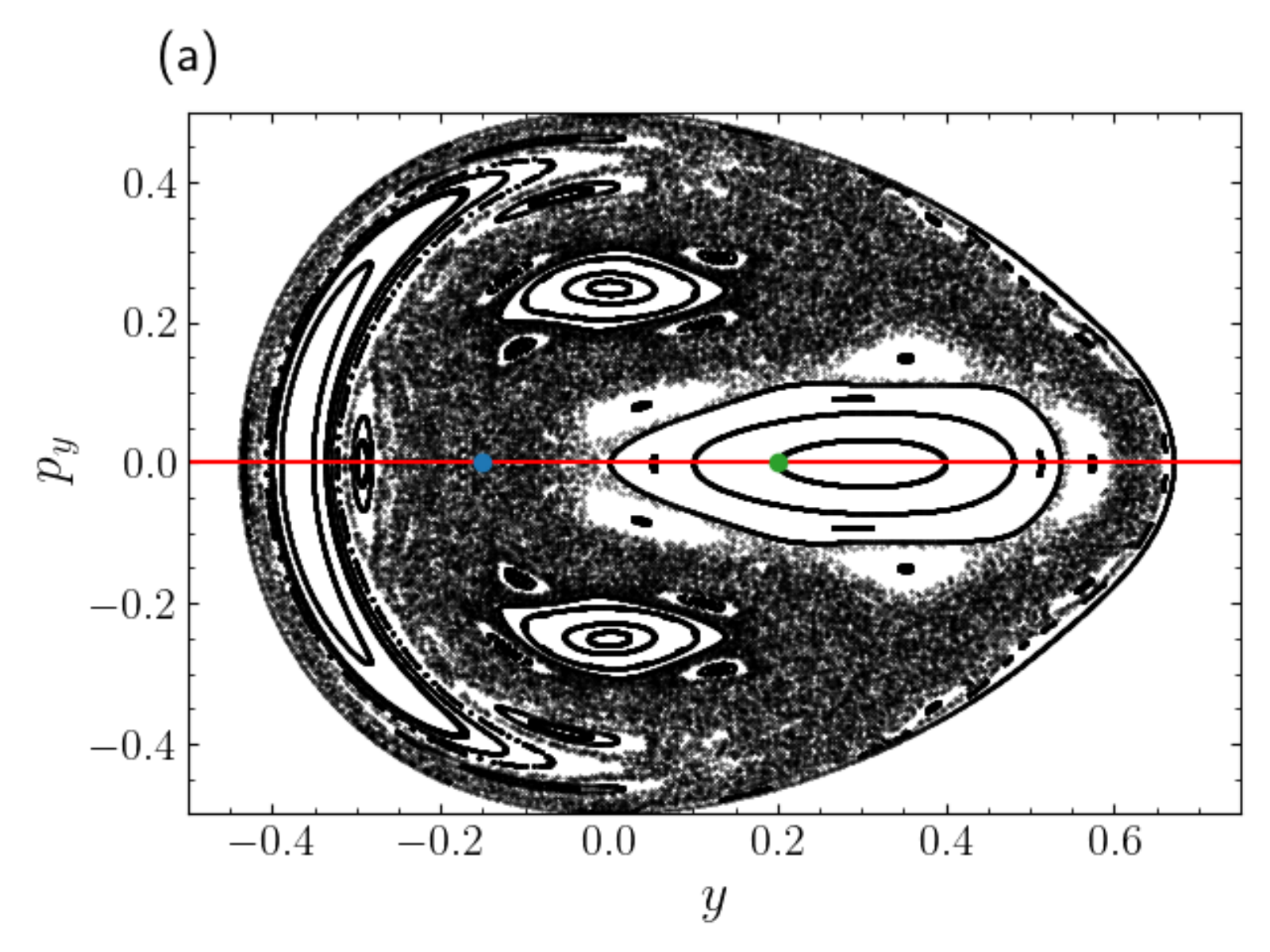}
    \includegraphics[width=0.9\columnwidth]{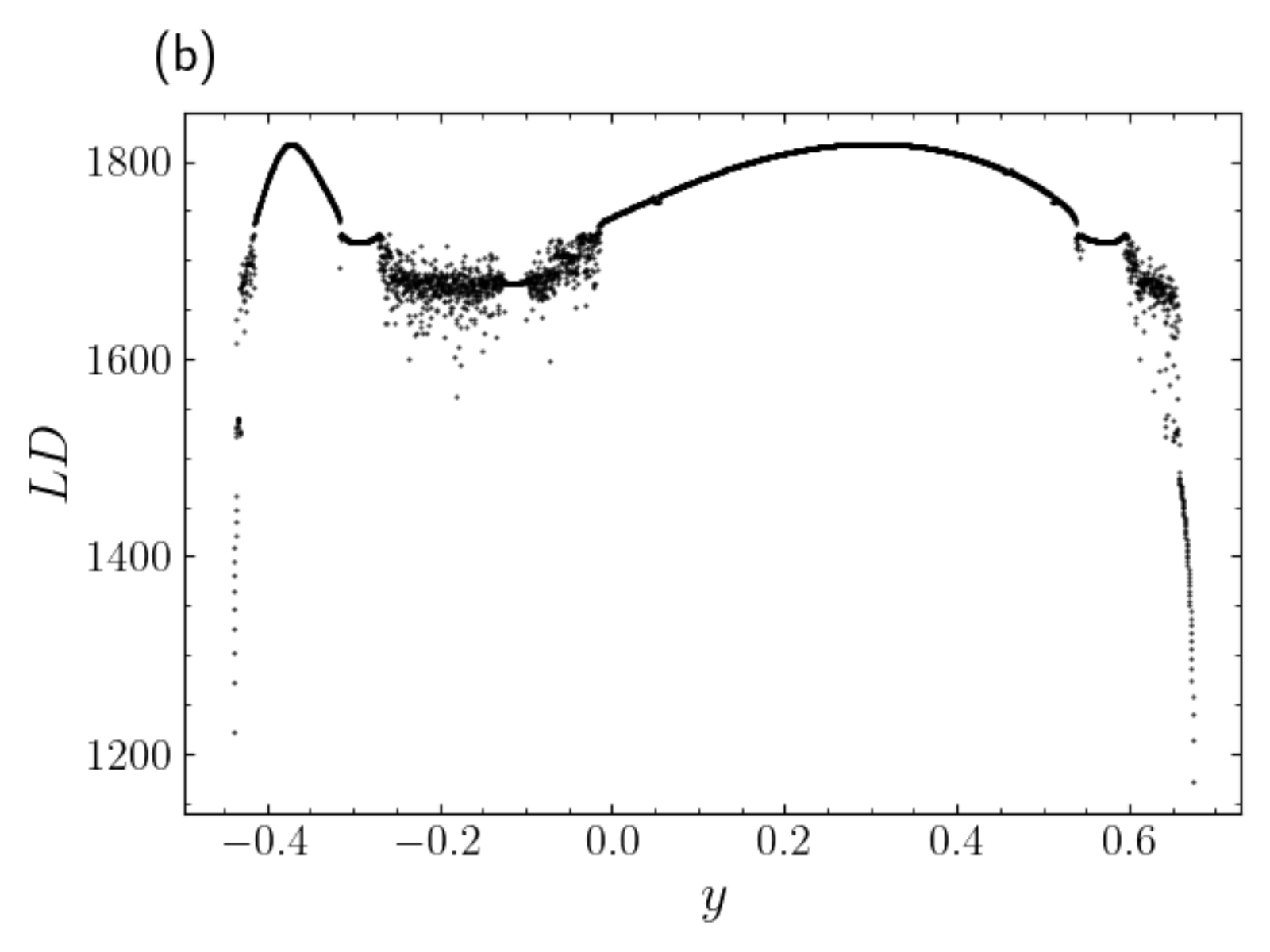}
    \includegraphics[width=0.9\columnwidth]{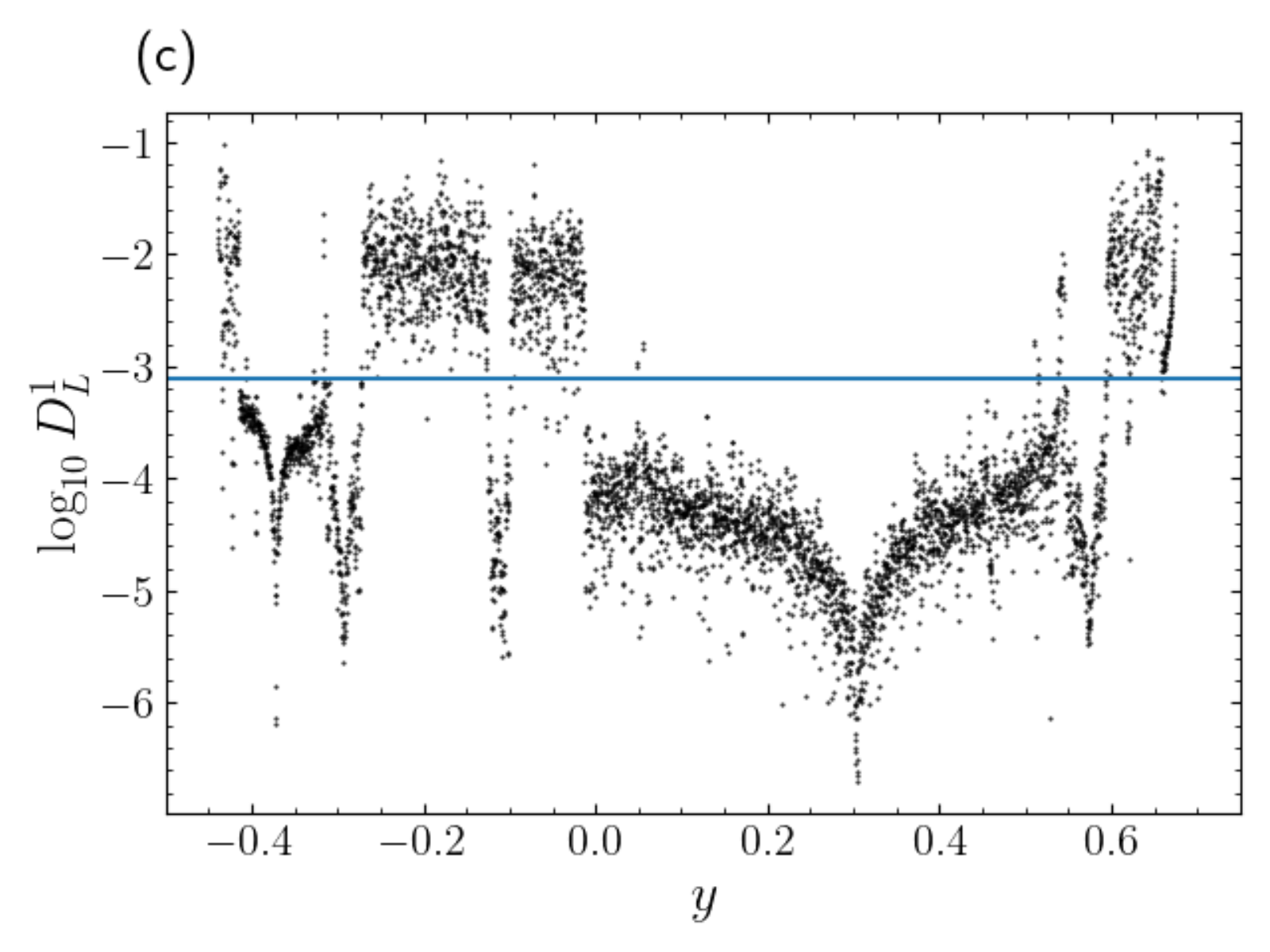}
    \includegraphics[width=0.9\columnwidth]{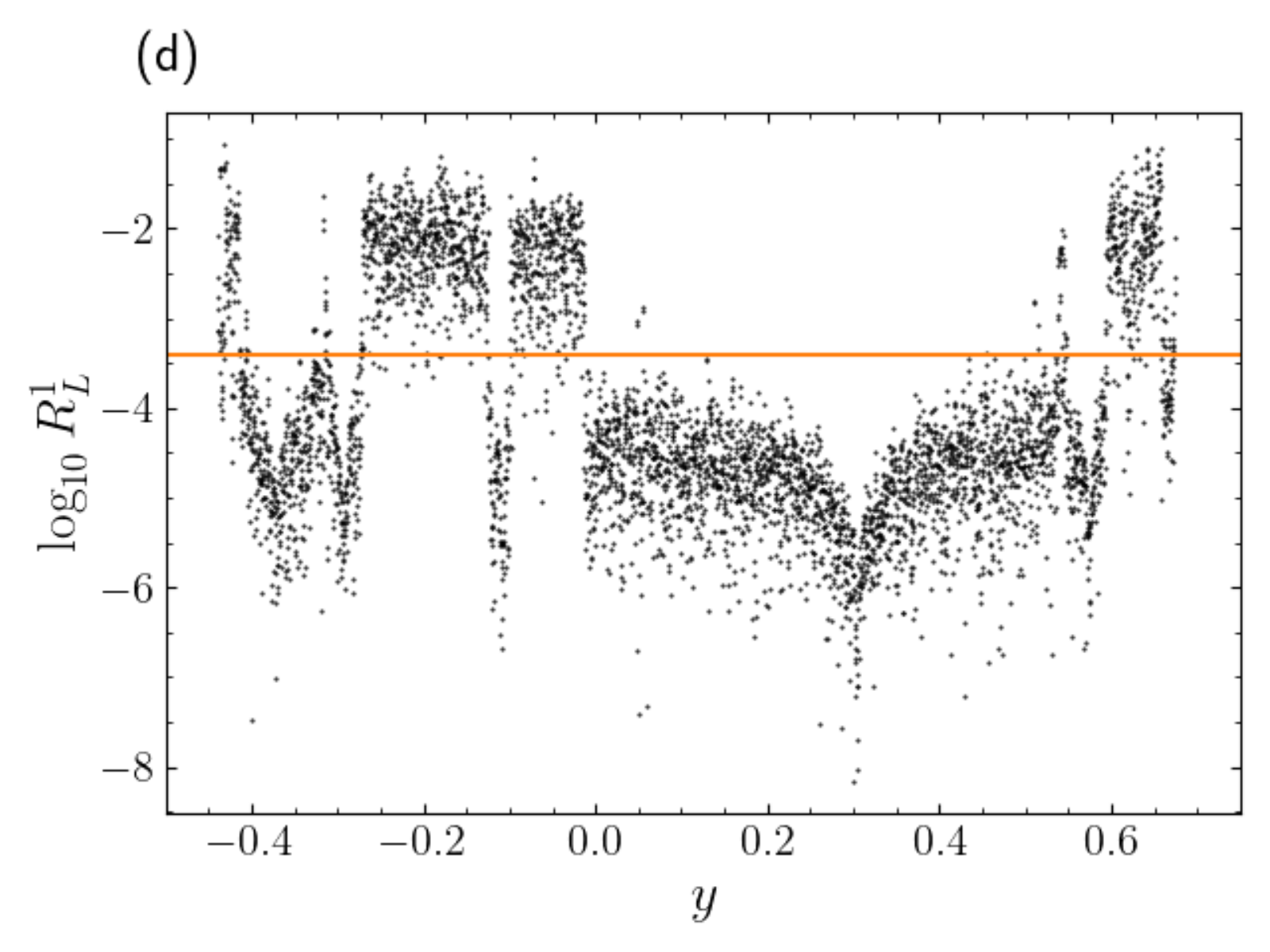}
    \includegraphics[width=0.9\columnwidth]{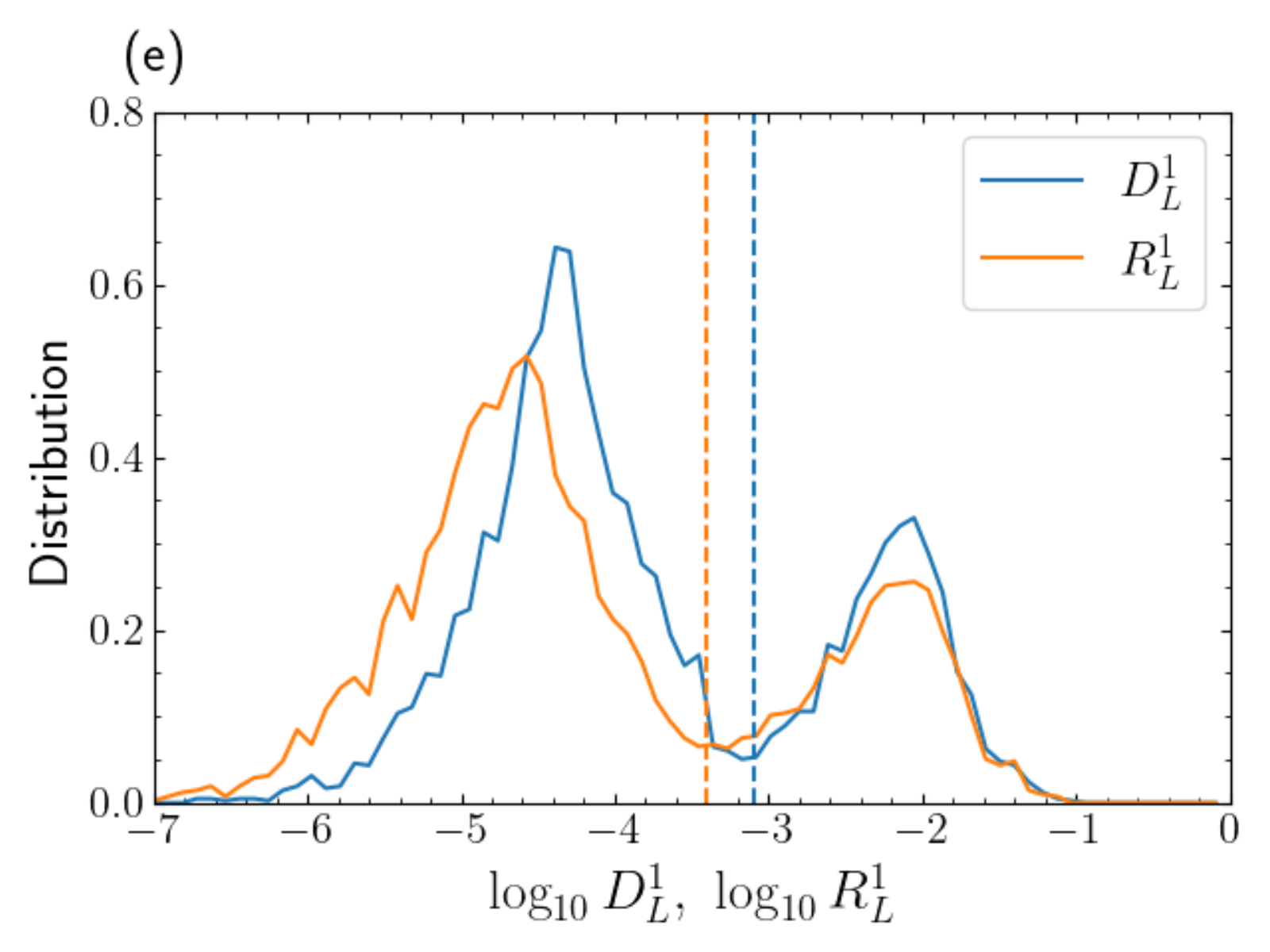}
    \includegraphics[width=0.9\columnwidth]{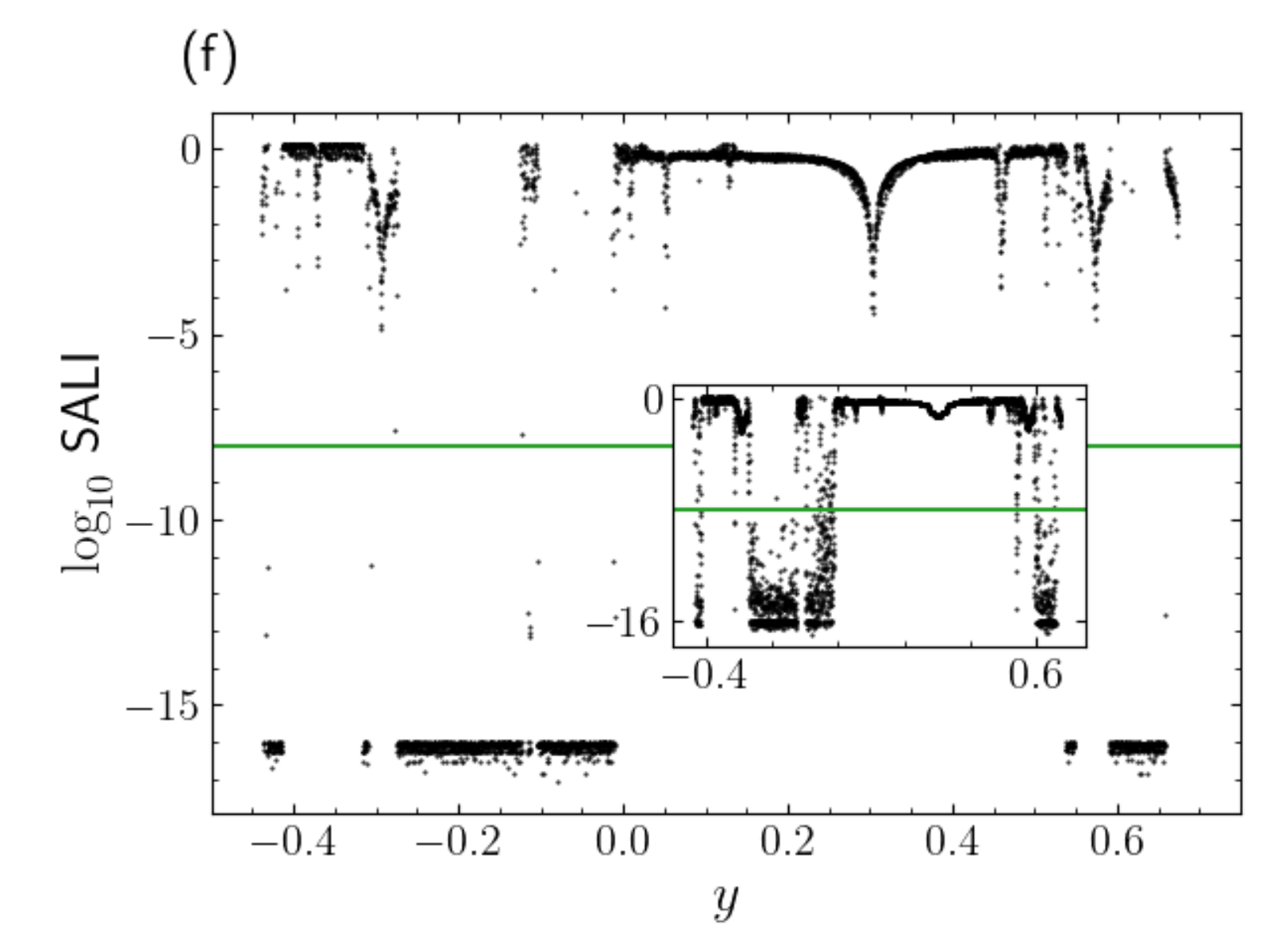}
      \caption{(a) The PSS (defined by $x=0$, $p_x>0$) of the H\'enon--Heiles system \eqref{eq:defHH} with energy $H=1/8$. The IC of the regular and chaotic orbit considered in Fig.~\ref{fig:NLD_illustrate} is respectively denoted by a blue and a green dot on the $p_y=0$ line (red horizontal line). (b) The forward LDs, at $\tau=10^3$, of about $4,500$ orbits whose ICs are homogeneously distributed  on the $p_y=0$ line  of the system's PSS [red horizontal line in panel (a)], as a function of the IC's $y$ coordinate. The values of $\log_{10} D_L^1$ (c), and  $\log_{10} R_L^1$ (d)  of the orbits considered in (b), with respect to the $y$ coordinate of the orbits' IC. The values $\log_{10} D_L^1 = -3.1$ and $\log_{10} R_L^1 = -3.4$ are respectively denoted by a horizontal blue line in (c) and a horizontal orange line in (d). (e) Normalized distributions of the  $\log_{10} D_L^1$ (blue curve) and $\log_{10} R_L^1$ (orange curve) values of panels (c) and (d). The values $\log_{10} D_L^1 = -3.1$ and $\log_{10} R_L^1 = -3.4$ are respectively denoted by a vertical blue and orange dashed line. (f) The values $\log_{10} \mbox{SALI}$ after $\tau=10^6$ (Inset: $\tau=10^3$) time units, of the orbits of panel (b), with respect to the $y$ coordinate of their ICs. In both the main and the inset panel of (f) the horizontal green line denotes the value $\log_{10} \mbox{SALI}=-8$.}
    \label{fig:1d_125_cha_reg}
\end{figure*}

As a first step towards investigating the behavior and the performance of the $D_L^n$   and  $R_L^n$ indicators,  we present in Fig.~\ref{fig:NLD_illustrate} their time evolution for two representative orbits of the H\'enon--Heiles system. In particular, for energy $H=1/8$, we consider a chaotic orbit with IC $x=p_y=0$, $y=-0.15$ and a regular one with IC $x=p_y=0$, $y=0.2$ [note that the fourth coordinate, $p_x>0$, of the IC of both orbits is computed from \eqref{eq:defHH}]. The ICs for each of these orbits are denoted by a blue (chaotic) and a green (regular) point on the system's PSS depicted in Fig.~\ref{fig:1d_125_cha_reg}(a).
For each orbit, the computation of the two indicators is based on the LDs of the orbits themselves and of two neighboring ones lying on the line $p_y=0$ in the system's PSS [red horizontal line in Fig.~\ref{fig:1d_125_cha_reg}(a)], having a difference $\sigma = 2.5 \cdot 10^{-4}$ in their $y$ coordinates. In this way we compute the $D_L^1$ \eqref{eq:defDNLD} and $R_L^1$ \eqref{eq:defRNLD} indices, but similar behaviors to the ones seen in Fig.~\ref{fig:NLD_illustrate} are observed when higher order indices with $n>1$ are computed. 

From  the results of Fig.~\ref{fig:NLD_illustrate} we see that in the case of the chaotic orbit the values of both the $D_L^1$ (dashed blue curve) and the $R_L^1$ (dashed green curve) indices remain generally well above $10^{-3}$, except from some short time intervals for which the LDs of the neighboring orbits are practically opposite, and consequently $R_L^1$ decreases. The occurrence of this arrangement of LDs for orbits neighboring a chaotic one, which can result in very small RNLD values (which typically are expected for regular orbits) should be borne in mind, as it can lead to the mis-characterization of chaotic orbits. Since the duration of the dips of the $R_L^1$ values in Fig.~\ref{fig:NLD_illustrate} is very short, the number of mischaracterized chaotic orbits is expected to be small whenever the index is used for a more global investigation of the dynamics of an ensemble of several orbits, and consequently it  should not affect the overall efficiency of the index. On the other hand, the $D_L^1$ (orange curve) and the $R_L^1$ (red curve) values for the considered regular orbit show a clearly distinct behavior, as they exhibit an oscillatory motif, remaining much smaller (as expected) than in the case of the chaotic orbit. It is worth noting that the $R_L^1$ is slowly increasing but nevertheless remains orders of magnitude smaller with respect to the values it attains for the chaotic orbit.

After gaining insight of the behavior of the DNLD and RNLD indices for individual orbits, we can gradually start using these indicators  to obtain a more general understanding of the dynamics of the H\'enon--Heiles system.  We begin our exploration by considering orbits with ICs on a line in the system's PSS. More specifically, we set the system's total energy to $H=1/8$ and create in Fig.~\ref{fig:1d_125_cha_reg}(a) the PSS $(y, p_y)$ defined by $x=0$, $p_x>0$. In this PSS we can clearly see regions of chaotic behavior (scattered points) and areas where regular motion occurs (smooth curves). We consider a set of several equidistant ICs on the $p_y=0$ line in this PSS [horizontal red line in Fig.~\ref{fig:1d_125_cha_reg}(a)] by taking a  spacing between neighboring ICs of $\sigma = 2.5 \cdot 10^{-4}$ in the interval $-0.5 \leq y \leq 0.75$. For each one of these orbits (in total about $4,500$ ICs were energetically permitted) we compute its forward LD. The results are plotted in Fig.~\ref{fig:1d_125_cha_reg}(b) as a function of the $y$ coordinate of the orbits' IC. Then, using these LDs we compute the corresponding $D_L^1$  \eqref{eq:defDNLD} and the $R_L^1$ \eqref{eq:defRNLD} indices for $\tau = 1,000$ and present their values in Figs.~\ref{fig:1d_125_cha_reg}(c) and (d) respectively. We note that results similar to the ones seen in Figs.~\ref{fig:1d_125_cha_reg}(c) and (d) are obtained if we use computations of only the backward LDs or both the forward and backward LDs. Thus,  restricting our study to using only the forward LD values provide the same dynamical information, decreasing at the same time the computational cost for finding the DNLD and  RNLD indices.

By contrasting the results of Fig.~\ref{fig:1d_125_cha_reg}(b), where the LDs of the orbits are plotted, with the location of these  orbits on the PSS of Fig.~\ref{fig:1d_125_cha_reg}(a), we observe a clear difference between the behavior of the LDs in regular and chaotic regions. The LDs of regular orbits have a fairly smooth variation with $y$, while in the chaotic regions the LDs behave erratically. This qualitative variation in the behavior of the LDs between regions of chaos and regularity has already been noted in Ref.~[\onlinecite{montes2021}], although it was not connected with the construction of a diagnostic which would allow the discrimination between the two cases. 

The definitions of both the $D_L^n$~\eqref{eq:defDNLD} and the $R_L^n$~\eqref{eq:defRNLD} indices imply that, in general, the smooth behavior of LDs for regular regions would result in smaller DNLD and RNLD values, with respect to the ones obtained for chaotic orbits where abrupt and erratic changes of LD values between nearby orbits are observed. This general trend is indeed evident in Figs.~\ref{fig:1d_125_cha_reg}(c) and (d) where the values of $D_L^1$  and the $R_L^1$ are respectively shown. The values of both indices appear to be in the vicinity of $\approx 10^{-2}$ for most chaotic orbits, and below an approximate, rough threshold value of $10^{-3}$ for regular ones. Nevertheless,  the distribution of the DNLD and RNLD values for the studied set of ICs can be utilized to determine a more accurate threshold value for distinguishing between chaotic and regular orbits. In Fig.~\ref{fig:1d_125_cha_reg}(e) we present the normalized distribution of the $D_L^1$ (blue curve) and the $R_L^1$ values (orange curve) which are respectively depicted in Figs.~\ref{fig:1d_125_cha_reg}(c) and (d). We see that both distributions have a similar shape, exhibiting two well defined peaks. One peak is localized at high values of the indices and corresponds to chaotic orbits, while the other is related to regular orbits and appears at lower $D_L^1$ and $R_L^1$ values. A good cutoff point separating the two peaks (and consequently between chaotic and regular orbits) can be taken as the index value ($D_L^1$ or $R_L^1$) for which the distribution attains its minimum between the two peaks. These threshold values approximately correspond to $\log_{10} \alpha_D = -3.1$ and $\log_{10} \alpha_R = -3.4$  for the $D_L^1$ and  $R_L^1$ respectively and are denoted by vertical dashed lines in  Fig.~\ref{fig:1d_125_cha_reg}(e). These two threshold values are also indicated by horizontal lines in Figs.~\ref{fig:1d_125_cha_reg}(c) and (d), where it can be clearly seen that they correctly capture the separation between high and low $D_L^1$ and the $R_L^1$ values. It is worth noting that the choice of an appropriate  threshold value $\alpha_D$ ($\alpha_R$) through the creation of the DNLD (RNLD) distribution depends on the particular studied system  and the considered set of orbits, which will determine the exact placement of the separation between the distribution's peaks. 

So, using the above defined $\alpha_D$ ($\alpha_R$) threshold value, we can characterize orbits as either regular or chaotic depending on whether their $D_L^1$ ($R_L^1$) values are below or above this threshold. In order to test the accuracy of this classification, we also compute the SALI (on the same  grid of ICs as we did for  $D_L^1$ and $R_L^1$) in order to identify orbits as regular or chaotic. In particular, we do this computation  for two final times, $\tau=10^3$ [inset of Fig.~\ref{fig:1d_125_cha_reg}(f)], which is the same time we used of computing the forward LDs and the related $D_L^1$ and $R_L^1$ indices [Figs.~\ref{fig:1d_125_cha_reg}(b), (c) and (d)], and  $\tau=10^6$ [main panel of Fig.~\ref{fig:1d_125_cha_reg}(f)] in order to accurately reveal the true nature of studied orbits. From the results of  Fig.~\ref{fig:1d_125_cha_reg}(f) and its inset we see that, in accordance to~\eqref{eq:SALI_HH}, the $\log_{10}$SALI attains large values for regular orbits, while for chaotic ones it goes to zero, reaching values close to the computer double-precision accuracy ($\log_{10}\mbox{SALI} \approx -16$) quite fast. This clear dichotomy between large (regular orbits) and very small (chaotic orbits) $\log_{10}\mbox{SALI}$ values is not that clear for $\tau=10^3$ as many orbits exhibit in-between values. These are mainly sticky orbits, located at the borders of stability islands, which need more time to reveal their chaotic nature.~\cite{skokos2004}

Following Ref.~[\onlinecite{skokos2004}] we use  the  value $\log_{10}\mbox{SALI} = -8$  as a threshold to  discriminate between regular ($\log_{10}\mbox{SALI} > -8$) and chaotic orbits ($\log_{10}\mbox{SALI} \leq -8$)  in our ensemble, and compare these results with the ones obtained by using the values of $D_L^1$ and $R_L^1$ and the thresholds  $\log_{10} \alpha_D = -3.1$ and $\log_{10} \alpha_R = -3.4$. Assuming that the SALI accurately reveals  the nature of the orbits we find that the percentage $P_A$ of correctly characterized orbits by using the $D_L^1$ and $R_L^1$ indices are respectively $P_A\approx 95.1 \%$ ($P_A\approx 95.4 \%$) and $P_A\approx 95.6 \%$ ($P_A\approx 95.2 \%$) when the SALI values for $\tau=10^6$ ($\tau=10^3$) are used. It is of particular importance to note that these percentages are very high, and do not significantly change when we use the more accurate characterization of orbits obtained from SALI values at $\tau=10^6$, which means that the $D_L^1$ and $R_L^1$ indices are not only capable of appropriately capturing the overall behavior of the ensemble, but can do that on short times.

By carefully analyzing Figs.~\ref{fig:1d_125_cha_reg}(a)-(d) we see that $D_L^1$ mainly fails to correctly identify the regular nature of orbits with ICs at both edges of the permitted range of  $y$-axis values (i.e.~for $y\lesssim -0.4$ and $y\gtrsim 0.65$), where LDs exhibit a smooth but very steep  gradient [Fig.~\ref{fig:1d_125_cha_reg}(b)]. In these regions, the summation of large absolute differences between LDs of neighboring orbits in \eqref{eq:defDNLD}, leads to large $D_L^1$ values, something which is naturally expected in chaotic regions due to the abrupt and erratic variations of LDs' values.
This problem is expected to be somehow mitigated when the $R_L^1$ index is computed, as one of the ratios $M(\mathbf{y}_i^+) / {M(\mathbf{x})}$ and  $M(\mathbf{y}_i^-) / {M(\mathbf{x})}$ appearing in \eqref{eq:defRNLD} would be larger than 1 with the other being smaller than 1. Thus, their sum will not highly deviate from 1, resulting in relatively small $R_L^1$  values, which in turn means that the orbit will be (correctly) characterized as regular. Comparing the points at the extreme right edges of Figs.~\ref{fig:1d_125_cha_reg}(c) and (d), we see that this expectation is indeed correct, as many (but not all) points show $R_L^1$ values below the threshold $\log_{10} \alpha_R = -3.4$ [orange horizontal line in Fig.~\ref{fig:1d_125_cha_reg}(d)], while the $D_L^1$ values of these ICs are above the threshold value   $\log_{10} \alpha_D = -3.1$ [blue horizontal line in Fig.~\ref{fig:1d_125_cha_reg}(d)]. On the other hand, the fact that local steep, monotonic gradients in the LD values of neighboring orbits lead to small $R_L^1$ [which was beneficial for the correct characterization of regular motion at the edges of the $y$ value range in Fig.~\ref{fig:1d_125_cha_reg}(b)], could also  appear at random places inside the chaotic region, resulting in relatively small $R_L^1$, which in turn will lead to the wrong characterization of chaotic motion as regular (recall here the brief dips of the green dashed curve in Fig.~\ref{fig:NLD_illustrate}). This can be seen for example in the case of the chaotic regions in the range $-0.3 \lesssim y \lesssim 0$ for which more data points are below the threshold line (wrongly denoting the corresponding orbits as regular) in Fig.~\ref{fig:1d_125_cha_reg}(d) than in  Fig.~\ref{fig:1d_125_cha_reg}(c). 

Furthermore, both the $D_L^1$ and the $R_L^1$ techniques are expected to face difficulties in correctly revealing the chaotic nature of sticky orbits. This is because both indices are based on computations of the forward LDs, whose values are defined by the whole history of the dynamics, which in turn is heavily dominated by the initial, long regularly behaving phase of the orbit's evolution. This issue could be addressed by considering longer integration  times for the computation of LDs. Of course, this approach will lead to more accurate results, but will also cause the loss of a basic advantage of the  $D_L^1$ and the $R_L^1$ indices, namely their ability to reliably capture the basic characteristics of the dynamics (within a remarkably small error of 5\% for the case of Fig.~\ref{fig:1d_125_cha_reg}) by performing computationally cheap short time, coarse-grid simulations.

\subsection{Global dynamics of the H\'enon--Heiles system} 
\label{sub:on_a_2d_surface_of_section}

After demonstrating the fundamental characteristics of the DNLD and RNLD indicators, and building our understanding as to how we can use these indices to distinguish between regular and chaotic motion for a set of orbits on a line of the PSS  of the H\'enon--Heiles system, we perform here an in depth investigation of the system's global dynamics for different energy values. 

\begin{figure*}
    \centering
    \includegraphics[width=0.99\textwidth]{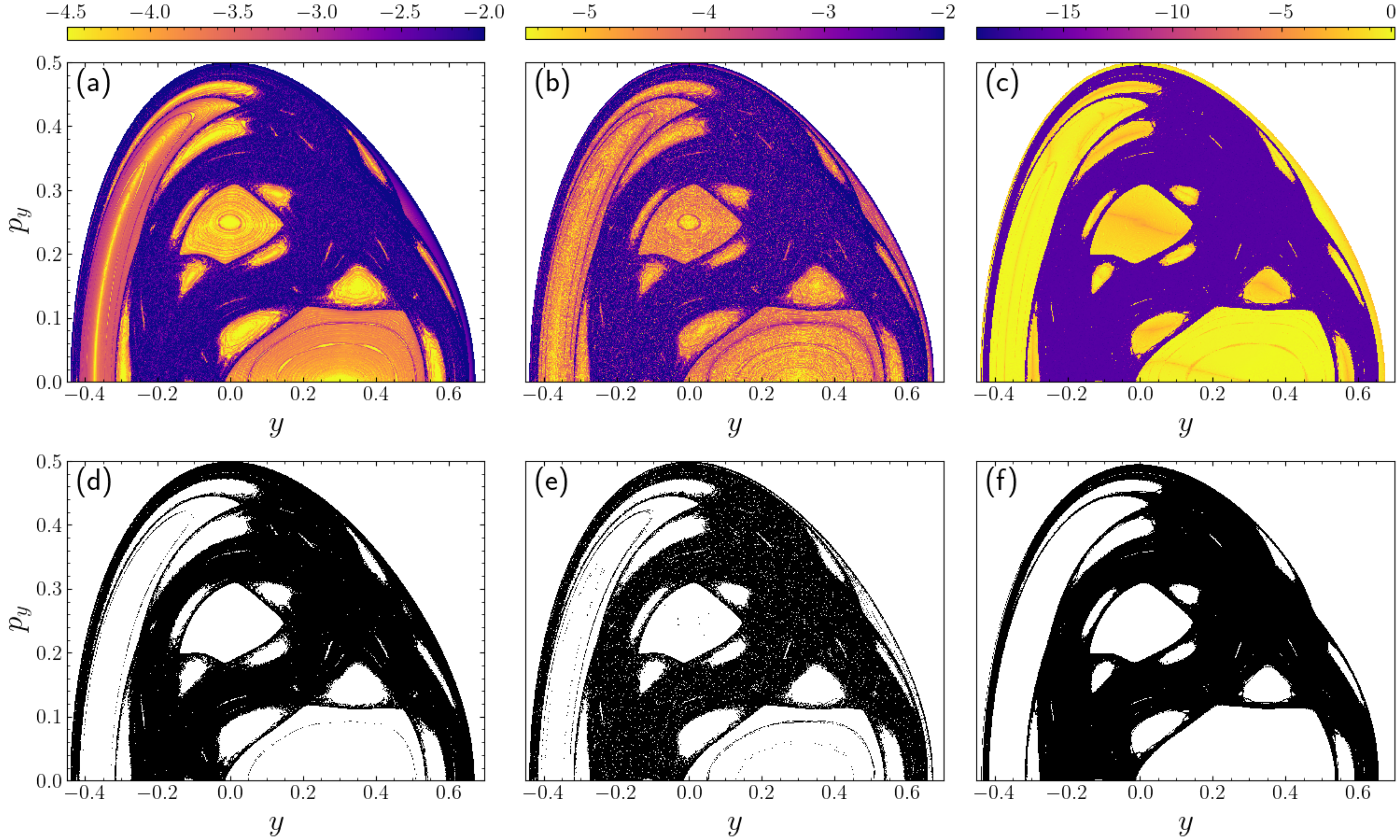}
   \caption{Results obtained  for orbits having their ICs on a $1,600 \times 800$ grid (around $865,000$ energetically permitted ICs) on the $p_y \geq 0$ part of the $x=0$, $p_x>0$, PSS of the H\'enon--Heiles system \eqref{eq:defHH} with energy $H=1/8$. The ICs are colored according to the orbits' (a) $\log_{10} D_L^2$,   (b) $\log_{10} R_L^2$, (c) $\log_{10} \mbox{SALI}$ values, using the color scales at the top of each panel. The results in (a) and (b) are based on computations of LDs for $\tau=10^3$ time units, while the SALI values in (c) are computed for $\tau=10^4$. Initial conditions of orbits characterized as chaotic by the (d) $D_L^2$ ($\log_{10} D_L^2 \geq -3$),   (e) $R_L^2$ ($\log_{10} R_L^2 \geq -3.4$), and (f) SALI ($\log_{10} \mbox{SALI} \leq -8$) indices.}
    \label{fig:2d_125_cha_reg}
\end{figure*}

First, we examine the case with energy $H=1/8$, which we already considered in Figs.~\ref{fig:NLD_illustrate} and \ref{fig:1d_125_cha_reg}. In order to reduce the required computational cost we make use of the fact that the PSS defined by $x=0$, $p_x>0$, is symmetric about the $p_y=0$ line [see Fig.~\ref{fig:1d_125_cha_reg}(a)], and restrict our investigation in its upper half specified by $p_y\geq 0$. In particular, we consider a grid of $1,600 \times 800$ equally spaced ICs in the region defined by $-0.5 \leq y \leq 0.75$ and $0 \leq p_y \leq 0.5$, setting in this way the distance between neighboring ICs to $\sigma=7.8125\cdot 10^{-4}$ in the $y$ direction and to $\sigma=6.25\cdot 10^{-4}$ in the $p_y$ direction. This arrangement leads to about $865,000$ energetically permitted ICs. We compute the LD of each one of these ICs for $\tau=1,000$ and then evaluate their $D_L^2$  \eqref{eq:defDNLD} and $R_L^2$ \eqref{eq:defRNLD} indices. It is worth noting here that since our ensemble of orbits lie on a $2$D subspace of the system's phase space, we implement the definitions of the DNLD $D_L^n$ \eqref{eq:defDNLD} and RNLD $R_L^n$ \eqref{eq:defRNLD} indices for $n=2$. The outcome of these computations are presented in Figs.~\ref{fig:2d_125_cha_reg}(a) and \ref{fig:2d_125_cha_reg}(b) where ICs are respectively colored according to their $\log_{10} D_L^2$  and $\log_{10} R_L^2$ value. By comparing these two color plots with the PSS of Fig.~\ref{fig:1d_125_cha_reg}(a) we clearly see that both indices manage to reveal the main features of the dynamics. In both Figs.~\ref{fig:2d_125_cha_reg}(a) and \ref{fig:2d_125_cha_reg}(b) islands of regular motion show up as regions of small  $\log_{10} D_L^2$ and $\log_{10} R_L^2$ values, while chaotic regions are represented by larger DNLD and RNLD values.

In order to further study the ability of the $D_L^2$  and the $R_L^2$ indices to correctly identify the regular or chaotic nature of orbits we first compute the SALI \eqref{eq:SALI} of each individual orbit to create a reference chart of regular (large $\log_{10}$SALI values) and chaotic (small $\log_{10}$SALI values) regions on the system's PSS. The computation of the SALI is done for $\tau=10^4$ time units, as this time is sufficient to create an accurate portrait of the dynamics at reasonable CPU times. The outcome of this process is shown in Fig.~\ref{fig:2d_125_cha_reg}(c). The direct comparison of this plot with Figs.~\ref{fig:2d_125_cha_reg}(a) and \ref{fig:2d_125_cha_reg}(b), shows the overall good diagnostic ability of the $D_L^2$  and  $R_L^2$ indices, as their implementation allows the identification of even small islands of stability in the chaotic sea and emphasizes the strength of these simple quantities for describing the phase space structure.

\begin{figure}
    \centering
    \includegraphics[width=0.99\columnwidth]{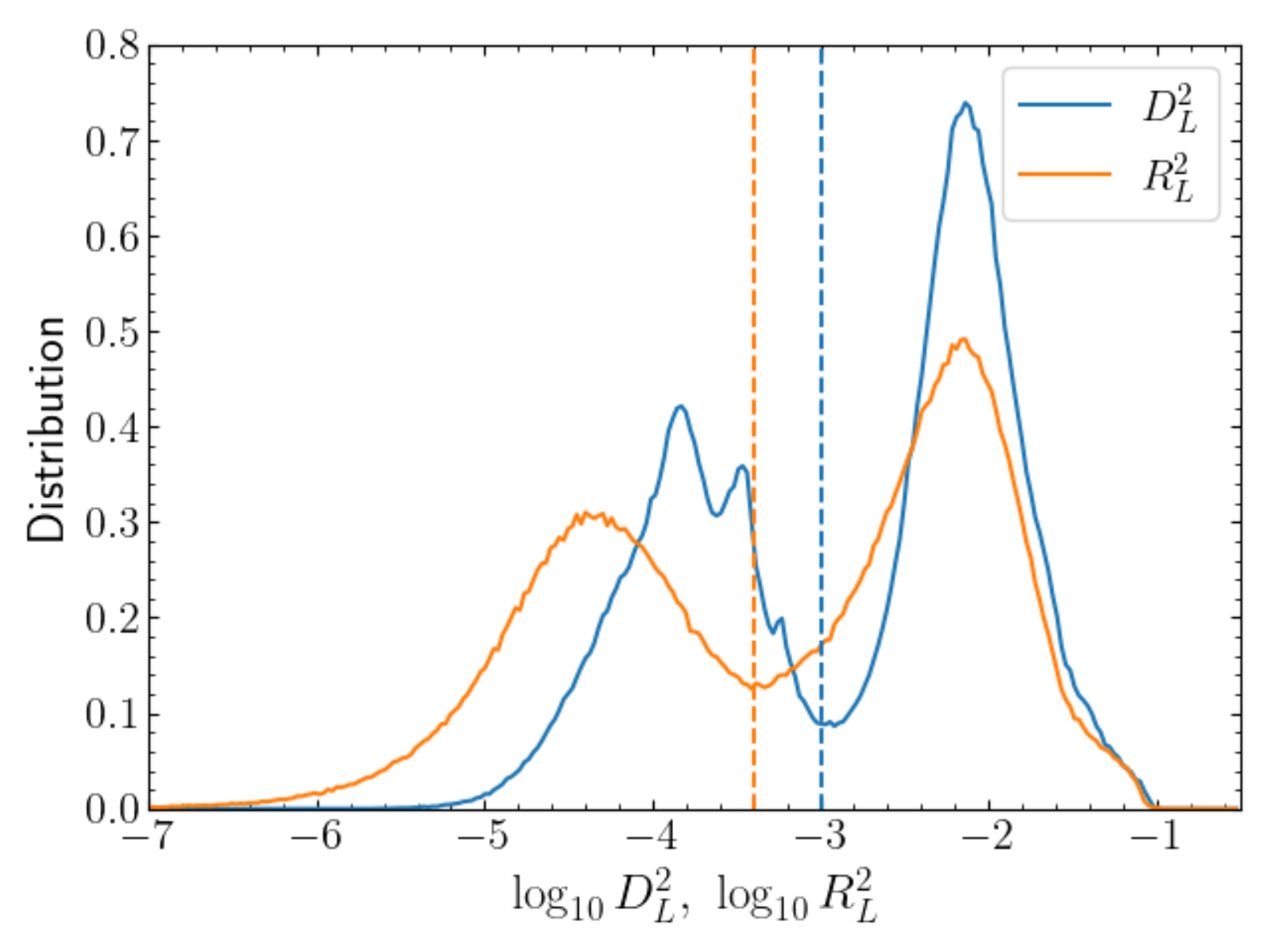}
    \caption{Normalized distributions of the  $\log_{10} D_L^2$ (blue curve) and $\log_{10} R_L^2$ (orange curve) values of the orbits considered in Figs.~\ref{fig:2d_125_cha_reg}(a)-(c). The values $\log_{10} D_L^2 = -3$ and $\log_{10} R_L^1 = -3.4$ are respectively denoted by a vertical blue and orange dashed line.} 
    \label{fig:2d_125_hists}
\end{figure}

Then, similarly to what was done for the orbits of Fig.~\ref{fig:1d_125_cha_reg}, we use the $D_L^2$  and  the $R_L^2$ values to characterize the orbit of each IC as regular or chaotic, and check the correctness of this characterization based on a similar analysis by exploiting the orbits' SALI values. In order to identify threshold values for the  $D_L^2$  and  $R_L^2$ indices to discriminate between regular and chaotic orbits we create in Fig.~\ref{fig:2d_125_hists} the normalized distributions of the computed  $\log_{10} D_L^2$ (blue curve) and $\log_{10} R_L^2$ (orange curve) values. It is worth noting that these distributions have similar shapes to the ones observed for the $D_L^1$  and $R_L^1$ indices in Fig.~\ref{fig:1d_125_cha_reg}(e). Then, as was done in that figure, we determine threshold values between the two peaks of each distribution, which will be used to separate the regular from the chaotic orbits. In this case we consider the following thresholds, $\log_{10} \alpha_D = -3$ and  $\log_{10} \alpha_R = -3.4$, which are respectively denoted by the blue and orange  dashed lines in  Fig.~\ref{fig:2d_125_hists}. Orbits with index values above this threshold are characterized as chaotic, with the remaining orbits identified as regular. In addition, as was done in Fig.~\ref{fig:1d_125_cha_reg}(f) orbits with $\log_{10}\mbox{SALI}\leq-8$ are labeled as chaotic, with regular orbits having   $\log_{10}\mbox{SALI} > -8$. 

Implementing these thresholds, we show in Figs.~\ref{fig:2d_125_cha_reg}(d)-(f) the ICs that are classified as chaotic by each indicator. The direct comparison of these figures show that both the $D_L^2$    [Fig.~\ref{fig:2d_125_cha_reg}(d)] and  the $R_L^2$ [Fig.~\ref{fig:2d_125_cha_reg}(e)] indices manage to correctly capture the structure of the chaotic component of the dynamics (and consequently the complementary regular part) as the obtained structures agree for the vast majority of points with the SALI classification [Fig.~\ref{fig:2d_125_cha_reg}(f)]. Nevertheless, some discrepancies can be identified. For example, we notice the existence of a few, thin regions inside some large stability islands,  which are incorrectly identified as chaotic by the $D_L^2$ and $R_L^2$ indices, in contrast to the classification provided by the SALI method. Furthermore, the difficulties faced by the indices (in particular the DNLD) when identifying regular behavior at the borders of the permitted PSS region, which was also present in the analysis of the results of Fig.~\ref{fig:1d_125_cha_reg}, is also evident here as for example the strip along the top right border of the PSS is incorrectly classified as chaotic. In addition, the density of points in the extended chaotic regions in Fig.~\ref{fig:2d_125_cha_reg}(e) is not very high as white points  are present, wrongly signifying the existence of regular orbits. This drawback of the RNLD index was also discussed in Sect.~\ref{sec:numerical_HH}.

\begin{figure}
    \centering
    \includegraphics[width=\columnwidth]{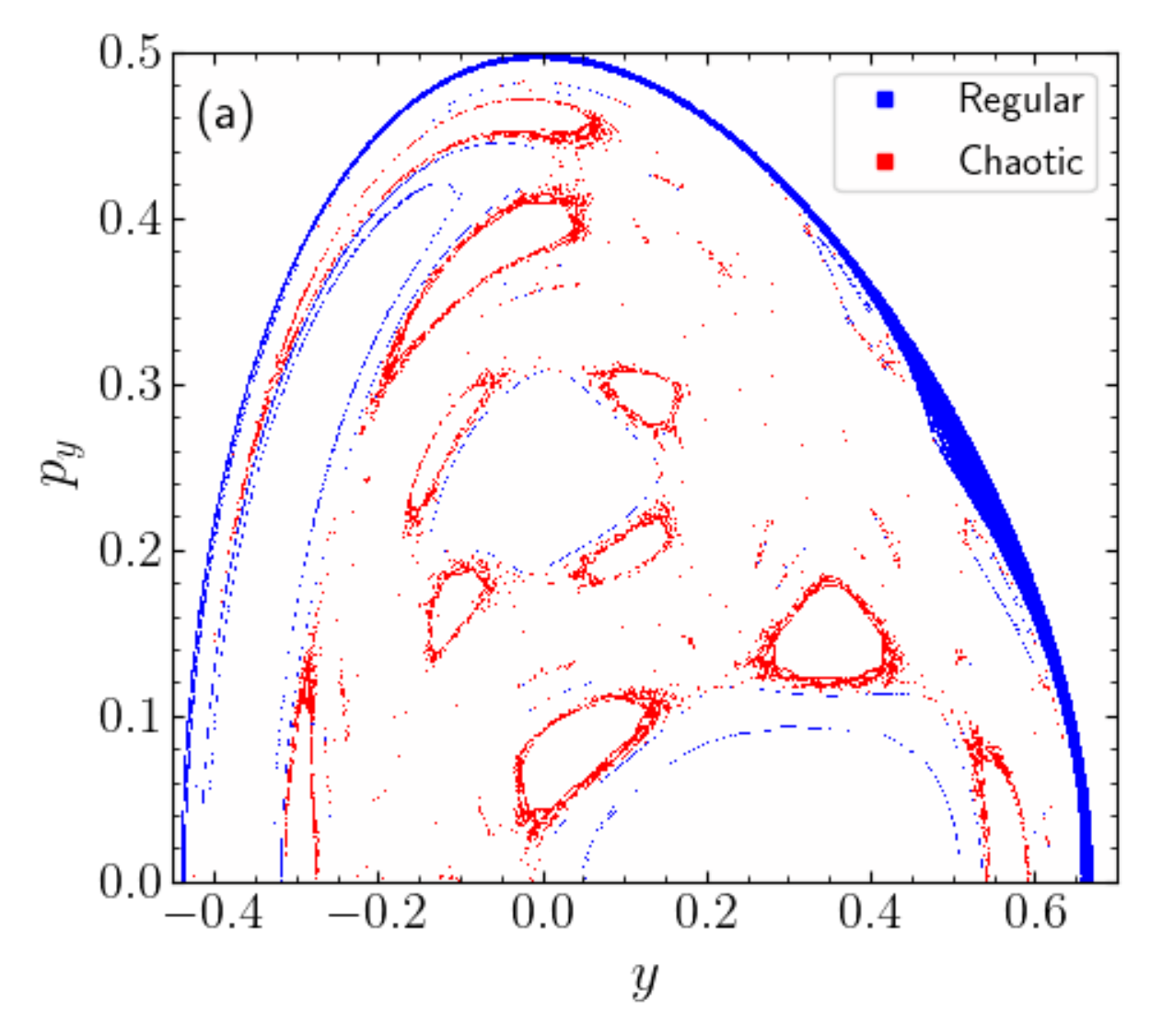} \\
    \includegraphics[width=\columnwidth]{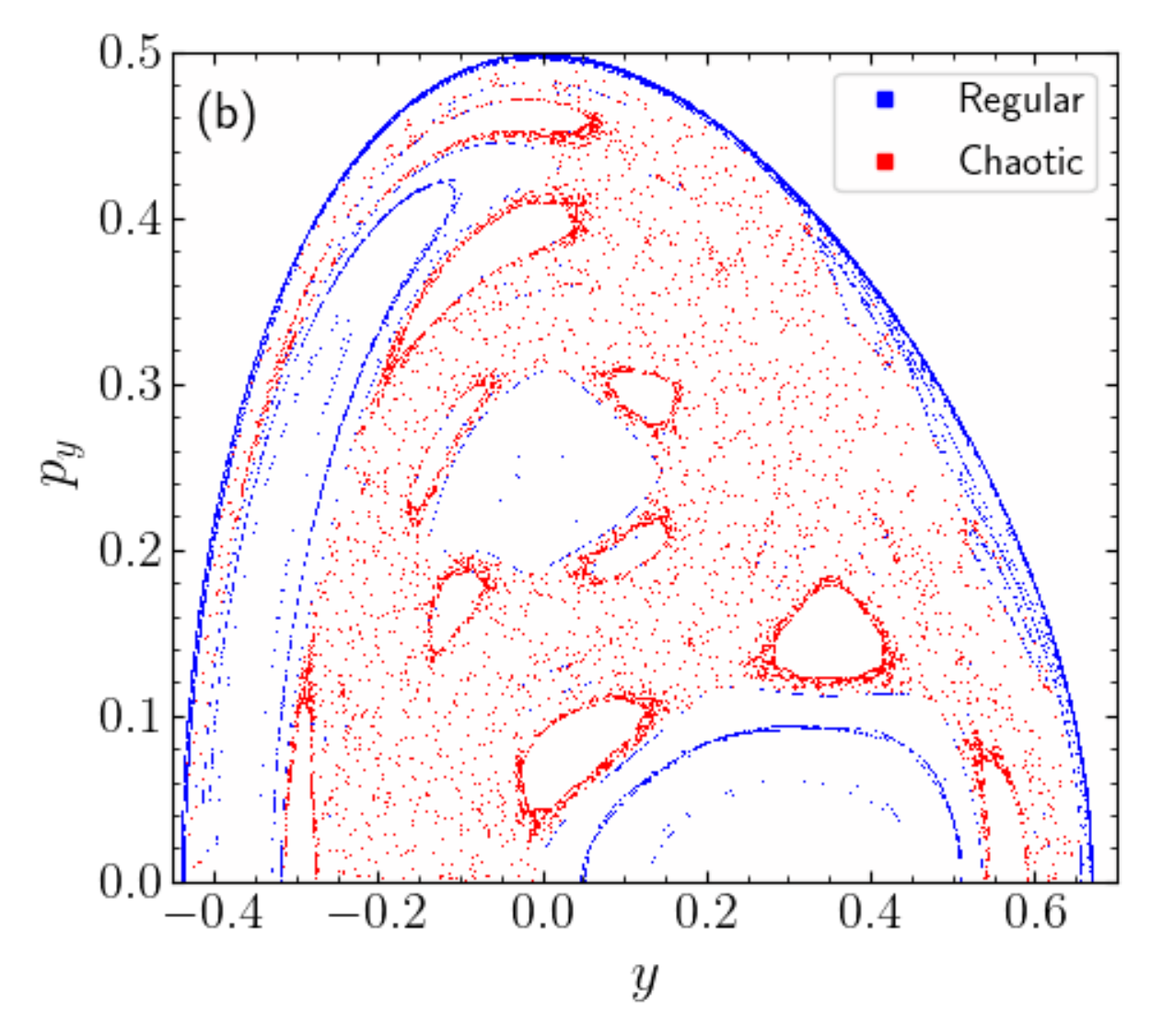}
    \caption{The set of ICs of Fig.~\ref{fig:2d_125_cha_reg} which are incorrectly characterized by the (a) $D_L^2$ and (b) $R_L^2$ index. In both panels blue points correspond to regular orbits (according to the classification obtained by the SALI method) which are falsely identified as chaotic, while red points denote  chaotic orbits which are incorrectly identified as regular.}
    \label{fig:2d_125_ic}
\end{figure}

In order to further substantiate and clearly reveal these discrepancies, we identify the ICs which are incorrectly characterized as chaotic or regular by the $D_L^2$ [Fig.~\ref{fig:2d_125_ic}(a)] and the $R_L^2$ [Fig.~\ref{fig:2d_125_ic}(b)] index, with respect to the identification obtained by the SALI method. More specifically, in Fig.~\ref{fig:2d_125_ic} we use blue (red) points for the ICs which, although the SALI identifies them as regular (chaotic) the $D_L^2$ [Fig.~\ref{fig:2d_125_ic}(a)] or the $R_L^2$ [Fig.~\ref{fig:2d_125_ic}(b)] index falsely characterizes them as chaotic (regular). From the results of Fig.~\ref{fig:2d_125_ic} it is easily seen that both indices fail to reveal the true nature of  sticky, chaotic orbits at the borders of stability islands, with the $D_L^2$ performing slightly worse than the $R_L^2$ as the thickness of the regions of red-colored points at the borders of stability islands is larger in Fig.~\ref{fig:2d_125_ic}(a) than in Fig.~\ref{fig:2d_125_ic}(b). On the other hand, the $R_L^2$ index incorrectly characterized more isolated chaotic orbits in the big chaotic sea of the system, as in Fig.~\ref{fig:2d_125_ic}(b) we observe more scattered red-colored points in that region. In addition, both indices have problems in revealing the regular nature of some orbits, and in particular the ones located  at the borders of the permitted PSS region, as the  presence of the blue-colored `layer' in that area denotes. It is worth noting  that the $D_L^2$ index performs slightly worse than the $R_L^2$ method in that region, as the larger width of the blue-colored `layer' indicates. Nevertheless, despite the incorrect identification of the nature of the ICs depicted in Fig.~\ref{fig:2d_125_ic}, both the $D_L^2$  and the $R_L^2$ indices manage to capture the overall behavior of the H\'enon--Heiles system \eqref{eq:defHH} for $H=1/8$, as is seen in Figs.~\ref{fig:2d_125_cha_reg}(d) and (e). In particular,  for the grid spacings and integration times used in Figs.~\ref{fig:2d_125_cha_reg} and \ref{fig:2d_125_ic}, we find that the characterization by the $D_L^2$ ($R_L^2$) of $P_A\approx 91.8 \%$ ($P_A\approx 92.3 \%$) of the orbits is in agreement with the results provided by the SALI method.

Based on the analysis of Figs.~\ref{fig:2d_125_cha_reg} and \ref{fig:2d_125_ic}, we expect the DNLD index to perform better for systems with large, extended chaotic regions, while the RNLD method would perform better in systems with larger regular regions. In order to investigate the validity of these predictions we perform a similar analysis to the one conducted in Figs.~\ref{fig:2d_125_cha_reg}, \ref{fig:2d_125_hists} and \ref{fig:2d_125_ic}, but for different energy values of the H\'enon--Heiles system, which result in different extents of the chaotic regions.

\begin{figure*}
    \centering
    \includegraphics[width=0.99\textwidth]{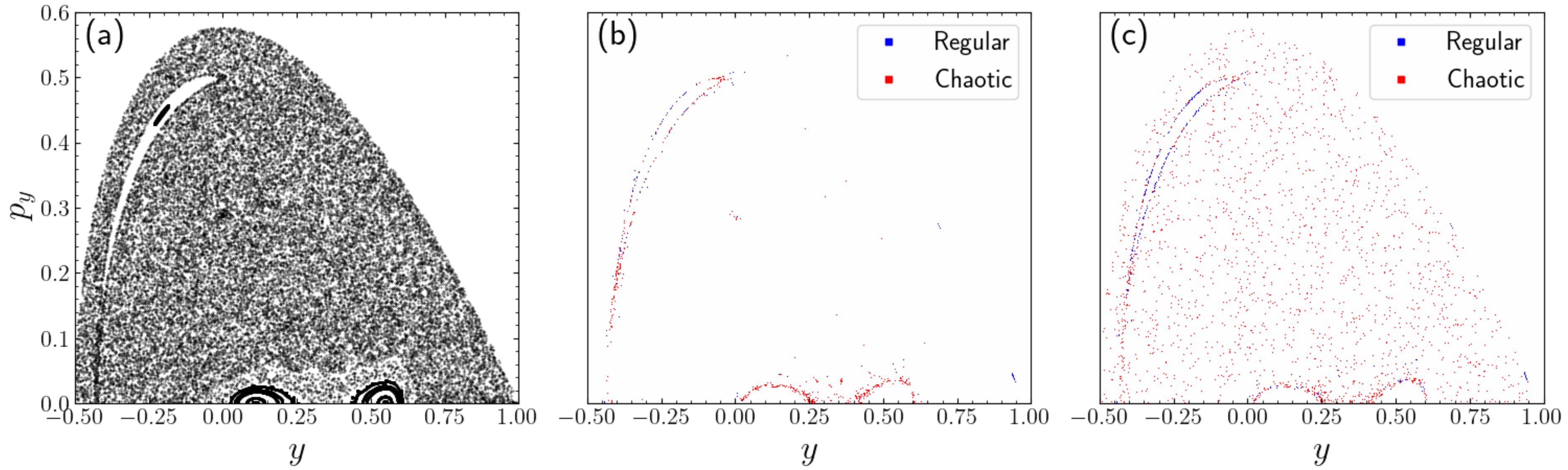}
    \caption{(a) The  $p_y \geq 0$ part of the $x=0$, $p_x>0$, PSS of the H\'enon--Heiles system \eqref{eq:defHH} with energy $H=1/6$.  The ICs  of orbits on the PSS of (a) which are incorrectly characterized by the $D_L^2$ and the $R_L^2$ index are respectively shown in (b) and (c) where  blue points correspond to regular orbits (according to the classification obtained by the SALI method) which are falsely identified as chaotic, while red points denote  chaotic orbits which are incorrectly identified as regular.}
    \label{fig:2d_166_cha_reg}
\end{figure*}

In Fig.~\ref{fig:2d_166_cha_reg} we present results for the H\'enon--Heiles system \eqref{eq:defHH} with energy $H=1/6$. From the system's PSS in Fig.~\ref{fig:2d_166_cha_reg}(a) we see the existence of a more extended chaotic region than was observed in  Fig.~\ref{fig:2d_125_cha_reg}(a) for $H=1/8$. Similarly to that case, we consider a grid of $1,600 \times 800$ equally spaced ICs in the region  $-0.5 \leq y \leq 1$ and $0 \leq p_y \leq 0.6$, so that  the distance between neighboring ICs is $\sigma=9.375\cdot 10^{-4}$ in the $y$ direction and $\sigma=7.5\cdot 10^{-4}$ in the $p_y$ direction. This setup yields about $852,000$ energetically permitted ICs. As before the evaluation of the $D_L^2$  and $R_L^2$ indices is based on computations of LDs for $\tau=10^3$, while the SALI values are computed for $\tau=10^4$. By performing a similar analysis to that presented in Fig.~\ref{fig:2d_125_hists} we set the threshold values for $D_L^2$  and $R_L^2$ to respectively $\log_{10} \alpha_D = -2.9$ and  $\log_{10} \alpha_R = -3.4$. Then, keeping the threshold value for SALI to $\log_{10}\mbox{SALI}=-8$, we obtain a $P_A\approx99.5 \%$ and $P_A\approx 98.0 \%$ agreement in the characterization of orbits when respectively the $D_L^2$  and the $R_L^2$ index is used in comparison to the SALI method. As predicted,  the $D_L^2$ index performs slightly better than the $R_L^2$ due to the fact that the system is mainly chaotic. The small number of incorrectly characterized ICs are mainly at the borders of the (few) stability islands for the  $D_L^2$ index [Fig.~\ref{fig:2d_166_cha_reg}(b)], while, as in Fig.~\ref{fig:2d_125_ic} for $H=1/8$, $R_L^2$ incorrectly characterizes isolated chaotic orbits in the extend chaotic sea. 

\begin{figure*}
    \centering
    \includegraphics[width=0.99\textwidth]{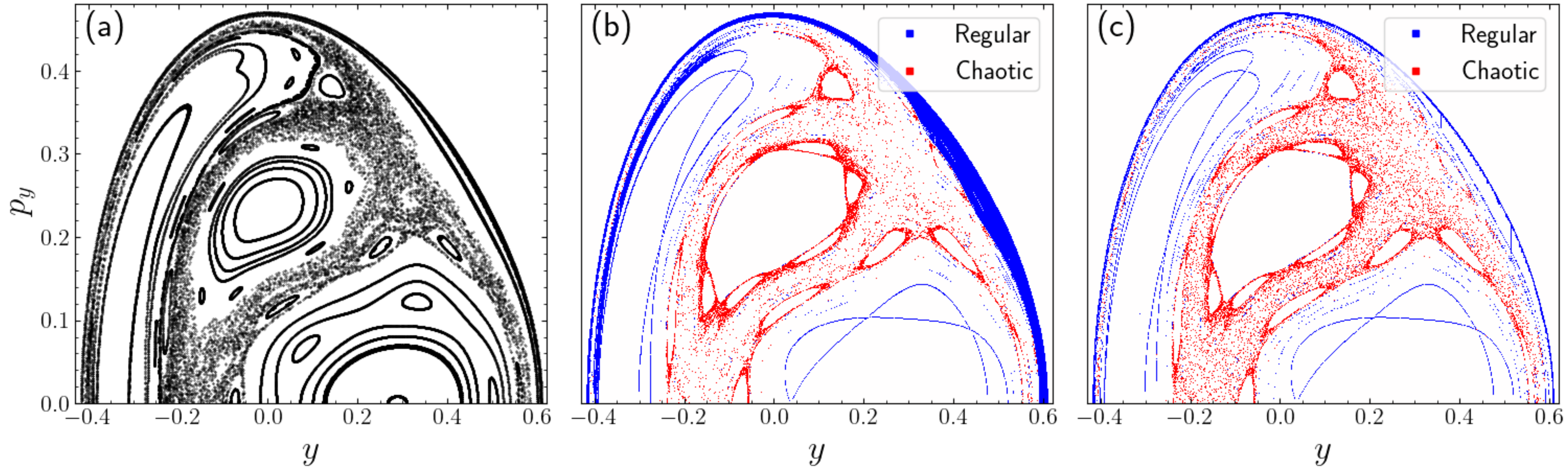}
    \caption{Similar to Fig.~\ref{fig:2d_166_cha_reg} but for the  H\'enon--Heiles system \eqref{eq:defHH} with energy $H=1/9$.}
    \label{fig:2d_111_cha}
\end{figure*}

In Fig.~\ref{fig:2d_111_cha} we present results similar to those in Fig.~\ref{fig:2d_166_cha_reg} but for $H=1/9$, which results in a PSS [Fig.~\ref{fig:2d_111_cha}(a)] with less chaos and correspondingly larger regular regions. Considering almost $760,000$ ICs on a $1,600 \times 800$ grid for $-0.5 \leq y \leq 1$ and $0 \leq p_y \leq 0.6$ (i.e.~$\sigma=7.8125\cdot 10^{-4}$ and $\sigma=6.25\cdot 10^{-4}$ in, respectively, the $y$ and $p_y$ direction), we compute, as in the previous cases,  the $D_L^2$, $R_L^2$ and SALI for each IC. Setting the threshold values for discriminating between regular and chaotic motion to  $\log_{10} \alpha_D = -3.1$, $\log_{10} \alpha_R = -3.3$, and  $\log_{10}\mbox{SALI}=-8$, we obtain an agreement of $P_A\approx 85.4 \%$ ($P_A\approx 88.5 \%$) between the $D_L^2$ ($R_L^2$) and the SALI orbit classification.  We show the initial conditions that are incorrectly characterized by the $D_L^2$ and the $R_L^2$ index in respectively  Figs.~\ref{fig:2d_111_cha}(b) and \ref{fig:2d_111_cha}(c). The large number of regular islands surrounded by sticky orbits at this energy results in the lower accuracy of both indices, with respect to their previous implementations. As predicted, the $R_L^2$ index performs better than the $D_L^2$ due to the increase of the regular component in the PSS. The borders of the energetically allowed phase space are  again (as in the $H=1/8$ case) incorrectly characterized mainly by the $D_L^2$ index [see the rather thick blue-colored region in Fig.~\ref{fig:2d_111_cha}(b)] resulting in a lower accuracy than exhibited by the $R_L^2$ index.

\subsection{Global dynamics of the 2D Standard map} 
\label{sub:a_2_dimensional_symplectic_map}

To further investigate the ability of the DNLD and the RNLD indices to reveal the chaotic behavior of discrete time dynamical systems, we apply them to the case of the well-known $2$D standard map, a symplectic mapping of the form~\cite{chirikov1979}
\begin{equation}
\label{eq:2Dmap}
\begin{array}{ll}
\displaystyle x_1' &= \displaystyle x_1 + x_2 ' \\
\displaystyle x_2' &= \displaystyle x_2 + \frac{K}{2 \pi}\sin\left(2 \pi x_1\right)
\end{array} \,\,\,\, (\mbox{mod} \, \, 1),
\end{equation}
with $K$ being a real parameter. Both variables are given (mod 1), so that $0 \leq x_i < 1$, $i=1,2$, while the prime $(\, ' \,)$ denotes the values of the coordinates after one iteration of the map.

\begin{figure*}
    \centering
    \includegraphics[width=0.99\textwidth]{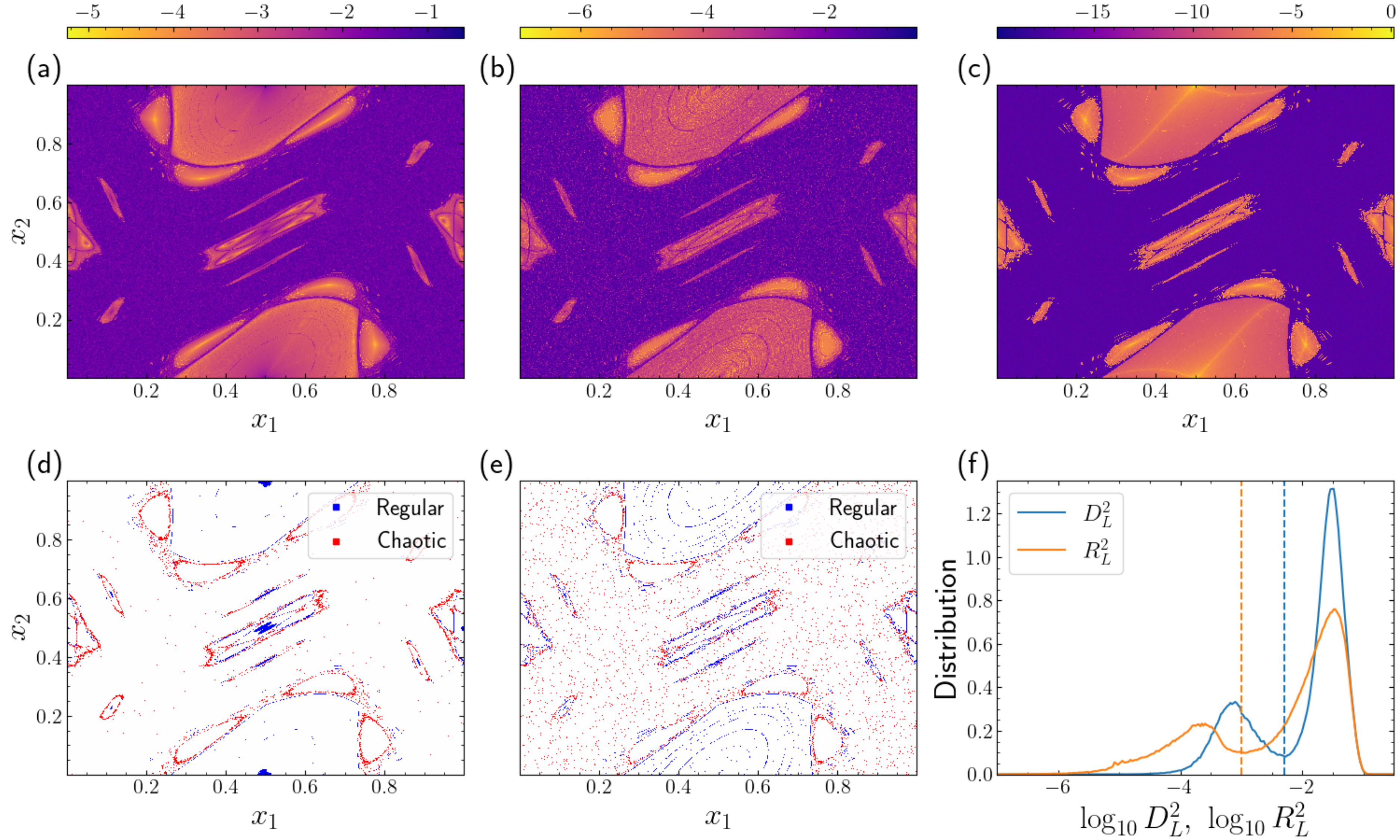}
    \caption{Results obtained for orbits having their ICs on a $1,200 \times 1,200$ grid on the phase space of the $2$D standard map \eqref{eq:2Dmap} with $K=1.5$. The ICs are colored according to the orbits' (a) $\log_{10} D_L^2$,   (b) $\log_{10} R_L^2$, (c) $\log_{10} \mbox{SALI}$ values, using the color scales at the top of each panel. The results in (a) and (b) are based on computations of LDs at $T=10^3$ iterations, while the SALI values in (c) are computed for $T=10^5$ iterations. The set of ICs of the map which are incorrectly characterized by the (d) $D_L^2$ and (e) $R_L^2$ index, with blue points corresponding to regular orbits (according to the classification obtained by the SALI method) which are falsely identified as chaotic, and red points denoting  chaotic orbits which are incorrectly identified as regular. (f) Normalized distributions of the computed $\log_{10} D_L^2$ (blue curve) and $\log_{10} R_L^2$ (orange curve). The values $\log_{10} D_L^2 = -2.3$ and $\log_{10} R_L^1 = -3$ are respectively denoted by a vertical blue and orange dashed line.}
    \label{fig:2dmap}
\end{figure*}

As a representative example of the system's dynamics we  consider  the case  $K=1.5$, for which the map  has a  large number of ICs  displaying regular or  chaotic behavior, and follow  the same steps of analysis as for the H\'enon--Heiles model. In particular, forward time LDs are computed on a square grid of $1,200\times1,200$ ICs for $T = 10^3$ iterations, and based on these values we evaluate the $D_L^2$ and the $R_L^2$ indices for each IC. In addition,  the  SALI values are computed on the same grid for a total number of $T=10^5$ iterations. The output of this process is seen in Fig.~\ref{fig:2dmap}. As in the case of the H\'enon--Heiles system [Figs.~\ref{fig:2d_125_cha_reg}(a)-(c)]  plots of the phase space where ICs are colored according to their $\log_{10} D_L^2$ [Fig.~\ref{fig:2dmap}(a)], $\log_{10} R_L^2$ [Fig.~\ref{fig:2dmap}(b)], and $\log_{10} \mbox{SALI}$ [Fig.~\ref{fig:2dmap}(c)] values reveal the same phase space structures.

Similarly to Figs.~\ref{fig:2d_125_ic}, \ref{fig:2d_166_cha_reg}(b)-(c) and \ref{fig:2d_111_cha}(b)-(c), we show in Figs.~\ref{fig:2dmap}(d) and \ref{fig:2dmap}(e) the ICs which are incorrectly characterized by the $D_L^2$ and the $R_L^2$ indices respectively, considering the identification obtained through their SALI values as true. The discrimination between regular and chaotic orbits by these indices is done by setting appropriate threshold values.  For the $D_L^2$ and the $R_L^2$ indices this is done through the construction of the distributions in  Fig.~\ref{fig:2dmap}(f), from which we respectively define the values $\log_{10} \alpha_D = -2.3$ and $\log_{10} \alpha_R = -3$ as appropriate thresholds separating the two peaks. Since according to \eqref{eq:SALI_2Dmap} the SALI is decreasing to zero for both regular (power law decrease) and chaotic (exponential decrease) orbits, an appropriate threshold value to distinguish between the two cases depends on the final number of iterations for which the index is computed. In our case, for which the SALI is evaluated after $T=10^5$ iterations, this threshold value is set to  $\log_{10}\mbox{SALI}=-12$. 

From the results of  Fig.~\ref{fig:2dmap}(d), we once more see that the $D_L^2$ index fails to correctly characterize regular regions of large but smooth LD gradients, like the ones which are now found in the centre of stability islands [see for example the blue colored area at the center of Fig.~\ref{fig:2dmap}(d)]. Furthermore, as we see from Fig.~\ref{fig:2dmap}(e), the $R_L^2$ index exhibits the same behavior we encountered for the  H\'enon--Heiles system  and mischaracterizes as regular several ICs in the large chaotic sea of the 2D map. Nevertheless, despite these discrepancies, both indices show a very good performance as they manage to correctly characterize  $P_A\approx 96.8 \%$ ($D_L^2$) and $P_A\approx 95.4 \%$ ($R_L^2$) ICs. These results clearly demonstrate that we can implement the DNLD and RNLD methods to reliably distinguish between regular and chaotic motion in low-dimensional conservative dynamical systems of both continuous autonomous Hamiltonians and area-preserving discrete time symplectic maps. 


\subsection{Effect of grid spacing and final computation time} 
\label{sub:effect_of_parameters}

Let us  now discuss in more detail the effect of  the final integration time ($\tau$) [or the total number of map iterations ($T$)], and  the grid spacing ($\sigma$) of neighboring ICs, on the performance of the DNLD and RNLD indices. 

It is important to note that the LDs themselves, and consequently the DNLD and RNLD indices, implicitly encode the exponential (polynomial) rate of divergence of nearby chaotic (regular) orbits. It is also reasonable to expect that very short integration times will be insufficient for a clear classification of chaotic and regular orbits, as in general, any numerical method requires a sufficient number of data to perform properly. More specifically, in our case we need long enough computations of LDs to clearly differentiate between exponential and polynomial growths, and so   very short numerical integrations are not expected to produce good results. Furthermore, as we discussed in Sect.~\ref{sec:numerical_HH}, since the evaluation of both the DNLD and RNLD indices is based on the whole evolution of orbits, very short time computations have difficulties in revealing the true nature of the dynamics, as for example we have repeatedly seen in cases of sticky orbits at the borders of stability islands. On the other hand, trying to create reliable short time diagnostics based on LDs is desirable to avoid unnecessarily long and CPU time consuming computations, especially since excessively long integration times can lead to an unclear pattern of behaviors. In order to understand the source of this drawback, we underline that the diagnostic power of the DNLD and RNLD methods resides in their ability to discriminate between the smooth and erratic variations of LD values of neighboring ICs [seen for example in Fig.~\ref{fig:1d_125_cha_reg}(b)], which are respectively encountered in regular and chaotic regions of the phase space. Thus, for very long times even the polynomial growth rates exhibited by regular orbits may be fast enough to create an apparently non-smooth behavior of LDs variations, which in turn will affect the diagnostic ability of the indices. 

With respect to the effect of the $\sigma$ on the performance of the DNLD and RNLD indices, we note that a very fine grid (i.e.~small distances between neighboring ICs) will require longer CPU times, as more orbits have to be integrated, but in principle should produce more accurate results. On the other hand, using a larger grid spacing will decrease the required computational cost but, at the same time, will inevitably lead to a less accurate description of the dynamics. Thus, a balance between these two aspects should be sought in every practical application of the indices.

\begin{figure}
    \centering
    \includegraphics[width=\columnwidth]{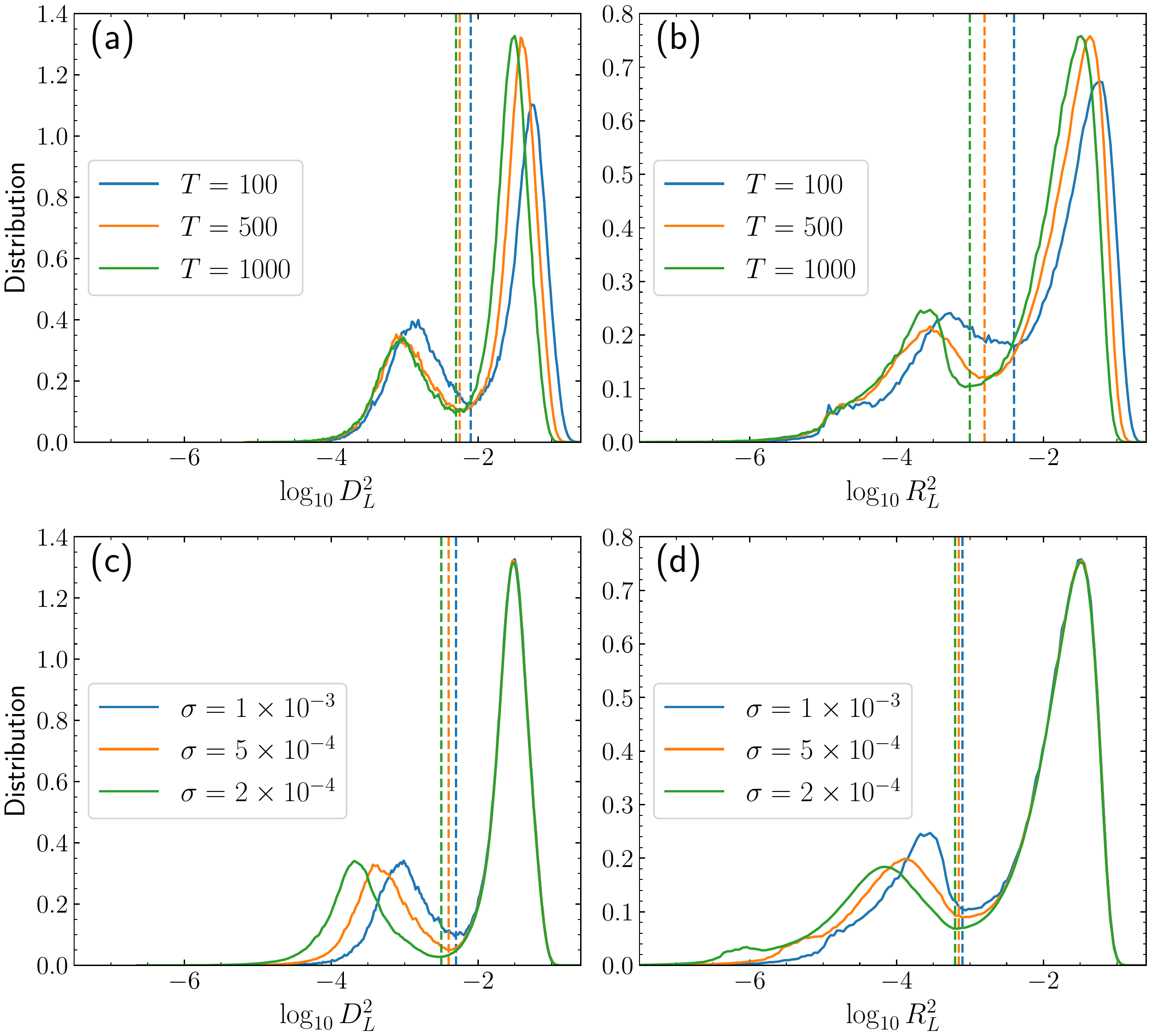}
    \caption{Normalized distributions of the [(a) and (c)]  $\log_{10}D^2_L$  and [(b) and (d)] $\log_{10}R^2_L$ values for the 2D standard map~\eqref{eq:2Dmap} with $K=1.5$. The results in (a) and (b) are obtained for a fixed spacing  $\sigma=10^{-3}$ and for three different  numbers of iterations $T$ of the map, whose explicit values are give in the legend. In (c) and (d) $T$ is fixed to $T=1,000$, while $\sigma$ varies. For each distribution, the threshold value chosen to separate the two peaks is shown by the respective dashed vertical line.}
    \label{fig:tau_sigma_distributions}
\end{figure}
\begin{figure*}
    \centering
    \includegraphics[width=0.99\textwidth]{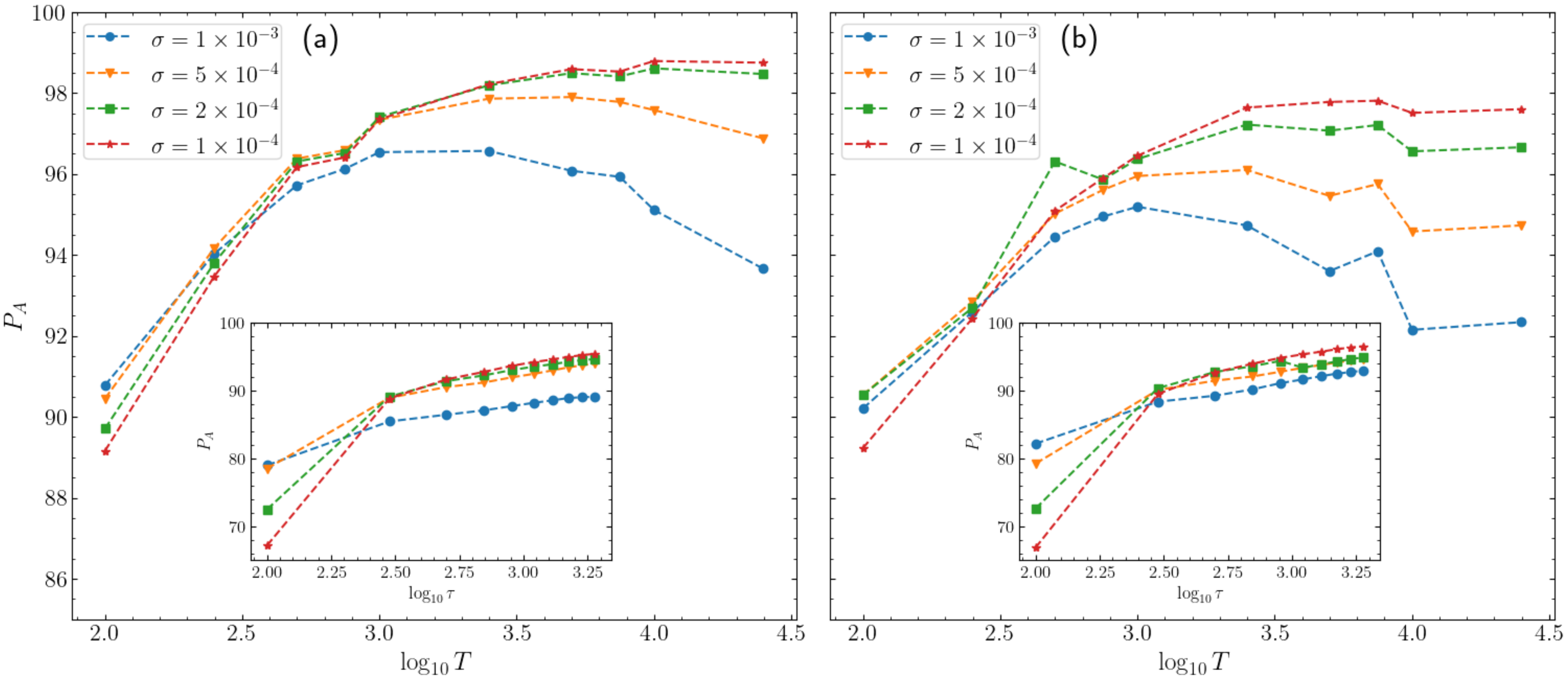}
    \caption{The percentage $P_A$ of orbits which are correctly  characterized by the (a) $D_L^2$ and (b) $R_L^2$ indices for the 2D map~\eqref{eq:2Dmap} with $K=1.5$  [insets: the H\'enon--Heiles system \eqref{eq:defHH} for $H=1/8$], with respect to the identification obtained by the SALI method for $T=10^5$ iterations [insets: $\tau=10^4$ time units], as a function of the number of iterations $T$ [insets: integration time $\tau$] used to compute the related LDs. Results are given for four different grid spacings $\sigma$, whose explicit values are reported in the legends and are the same for both systems. In the case of the 2D map (main panels) ICs were taken in the system's entire phase space with equal spacing $\sigma$ in both the $x_1$ and $x_2$ axes, while for the H\'enon--Heiles system (insets) the considered ICs were located in the $p_y \geq 0$ part of the system's PSS defined by $x=0$, $p_x>0$,   with the same spacing $\sigma$  used for both the $y$ and the $p_y$ coordinates.}
    \label{fig:accuracies}
\end{figure*}
It is worth noting that although in all considered cases in this study the DNLD and RNLD distributions retain their shape exhibiting two well defined peaks, similar to what is seen in Figs.~\ref{fig:1d_125_cha_reg}(e), \ref{fig:2d_125_hists} and \ref{fig:2dmap}(f), the exact value of the threshold used to distinguish between regular and chaotic orbits (defined to be the minimum between the two peaks) varies for  each $\sigma$ value and total number of iterations $T$. This behavior is, for example,  demonstrated in Fig.~\ref{fig:tau_sigma_distributions}, where we show the distributions of the $\log_{10}D^2_L$ [Figs.~\ref{fig:tau_sigma_distributions}(a) and \ref{fig:tau_sigma_distributions}(c)] and the  $\log_{10}R^2_L$  [Figs.~\ref{fig:tau_sigma_distributions}(b) and \ref{fig:tau_sigma_distributions}(d)] values for the 2D standard map~\eqref{eq:2Dmap} with $K=1.5$ for various $\sigma$ and $T$ values. The distributions in Figs.~\ref{fig:tau_sigma_distributions}(a) and \ref{fig:tau_sigma_distributions}(b) are obtained for $\sigma=10^{-3}$ and for three different $T$ values, namely $T=100$ (blue curves),  $T=500$ (orange curves) and $T=1,000$ (green curves). The vertical dashed lines indicate the threshold values near the minimum of these distributions, which are used to discriminate between chaotic and regular orbits. Here we see a slight decrease in the threshold value with increasing $T$, which is more  evident for the $R^2_L$ index [Fig.~\ref{fig:tau_sigma_distributions}(b)]. Similarly, in Figs.~\ref{fig:tau_sigma_distributions}(c) and \ref{fig:tau_sigma_distributions}(d) we respectively present the distributions of the  $\log_{10}D^2_L$ and the $\log_{10}R^2_L$ values obtained for $T=1,000$ and $\sigma=10^{-3}$ (blue curves), $\sigma=5\cdot 10^{-4}$ (orange curves) and $\sigma=2\cdot 10^{-4}$ (green curves). From the results of these figures it is seen that a varying grid size does not significantly change the position of the threshold for both indices, although a slight decrease of the threshold value with decreasing $\sigma$ is observed. 

In the main panels of Fig.~\ref{fig:accuracies} we show, for the 2D standard map \eqref{eq:2Dmap} with $K=1.5$, how the percentage $P_A$ of ICs correctly characterized by the $D_L^2$~[Fig.\ref{fig:accuracies}(a)] and $R_L^2$~[Fig.~\ref{fig:accuracies}(b)] methods (when compared with SALI computations for $T=10^5$ iterations) changes with the number of iterations $T$ used to compute the related LDs, for several $\sigma$ values defining a symmetric grid in the phase space of the system.
For both indices the $P_A$ values initially increase as $T$ grows for all $\sigma$ values, showing a rise of at least $5\%$ when $T$ changes from $T=10^2$ to  $T=10^3$. After that point, for the coarser grids with $\sigma=10^{-3}$ (blue circles in Fig.~\ref{fig:accuracies}) and $\sigma=5\cdot 10^{-4}$ (orange triangles in Fig.~\ref{fig:accuracies}), we see a decrease in performance, which is likely related to the predicted introduction of confusion in regular regions where the polynomial divergence of nearby orbits could lead to non-smooth LD variations. It is also evident that, after some $\sigma$ value, the further decrease of $\sigma$ does not significantly change the behavior of $P_A$ for increasing $T$ values, as the results for $\sigma=2\cdot 10^{-4}$ (green squares in Fig.~\ref{fig:accuracies}) and $\sigma=1\cdot 10^{-4}$ (red stars in Fig.~\ref{fig:accuracies}) are rather similar. The insets in each panel of Fig.~\ref{fig:accuracies} show  analogous results to the main panels, but for the PSS (defined by $x=0$, $p_x>0$) of the H\'enon--Heiles Hamiltonian \eqref{eq:SALI_HH} with $H=1/8$, when the SALI is computed at $\tau=10^5$ time units. Also here we see a  drastic initial improvement with increasing $\tau$ values, followed by a more moderated one, but the turning point to lower $P_A$ values does not appear at the timescales studied.

The main message of this analysis is that we have to balance the desired accuracy against the required computational time in numerical applications of the DNLD and RNLD indices.
It is also very clear from the results of Fig.~\ref{fig:accuracies} that simply increasing the LD computation time very quickly loses value as an approach to improve accuracy. A sufficiently high $T$ (or $\tau$) value is needed for adequately capturing  the basic features of the dynamics, but the further increase of the final integration time requires a fairly fine grid to be useful, and only provides marginal gains as the associated computational cost increases significantly. From Fig.~\ref{fig:accuracies} we see that for a mixed phase space, with relatively large chaotic and regular regions, it is eminently possible to obtain an accuracy of up to $P_A\approx 95\%$ with the DNLD and RNLD methods, using feasible grid sizes and relative short final integration times.

Finally, we note that the CPU time required for identifying the ensembles of orbits considered in our study as regular or chaotic by the DNLD and RNLD indices based on short time LD computations is typically three times smaller that the performed SALI computations for the same values of $\tau$ (or $T$). The reason for this difference is the fact that the SALI computation for each IC requires the simultaneous integration of two deviation vectors, apart from the orbit itself, while the  DNLD and RNLD indices circumvent this requirement by using the computed LD values on the grid points. Of course the SALI provides more accurate results, but reaching accuracy levels of $P_A \gtrsim 90\%$ with the DNLD and RNLD indices is a very useful alternative, which allows us to have a reliable overall understanding of the system's global behavior at a lower computational cost. This alternative will be especially usefully in cases where the variational equations are inaccessible, or very difficult to  obtain, or sufficiently complicated that their numerical integration will require large CPU times.

\section{Summary and discussion} 
\label{sec:conclusions}

We have introduced and successfully implemented computationally efficient ways to effectively identify chaos in low-dimensional conservative dynamical systems from the values of LDs at neighboring ICs. The conceptualization of these methods is based on the observation that LDs show a smooth variation with regard to ICs in phase space regions where regular motion occurs, in contrast to an erratic behavior seen in chaotic regions. More specifically, we introduced two indices which manage to quantify these changes in the behavior of LDs of nearby ICs: The Difference of Neighboring orbits’ Lagrangian Descriptors (DNLD), $D^n_L$ \eqref{eq:defDNLD}, and the  Ratio of Neighboring orbits’ Lagrangian Descriptors (RNLD), $R^n_L$ \eqref{eq:defRNLD}. These indicators use the information about the rate of divergence of nearby ICs encoded into a precomputed grid of LD values to accurately estimate the chaoticity of orbits. 

By performing several numerical simulations, we confirmed that both quantities are eminently practical for detecting regions of chaotic and regular motion in both the 2dof autonomous H\'enon--Heiles Hamiltonian system \eqref{eq:defHH} and the area-preserving 2D standard map~\eqref{eq:2Dmap}, which are widely used  prototypical models of low-dimensional conservative systems, as they exhibit all the basic dynamical features appearing in such systems.
In particular, the creation of color plots of the systems' PSS [Figs.~\ref{fig:2d_125_cha_reg}(a) and \ref{fig:2d_125_cha_reg}(b)] or phase space [Figs.~\ref{fig:2dmap}(a) and \ref{fig:2dmap}(b)], where ICs are colored according to the corresponding DNLD or RNLD value, clearly show that both indices manage to reveal the main characteristics of the dynamics. 

Apart from this  qualitative feature we investigated in detail the ability of the indices to provide reliable, quantitative results about the chaoticity of the studied systems. More specifically, we used the distributions of the indices' values to determine appropriate threshold values which allow the characterization of orbits as regular or chaotic. For all studied cases, for which regular and chaotic orbits coexist, the DNLD and RNLD distributions have similar shapes exhibiting two well defined peaks [Figs.~\ref{fig:1d_125_cha_reg}(e), \ref{fig:2d_125_hists}, \ref{fig:2dmap}(f) and \ref{fig:tau_sigma_distributions}. Then the value of the index for which the distribution shows its minimum between the two peaks is used as the threshold value discriminating between chaotic and regular orbits. This process does not lead to a universal outcome, as the obtained threshold values vary for  different final integration times $\tau$ (number of iterations $T$) and distances $\sigma$ between neighboring orbits. We implemented this approach and compared our classification with that obtained by using a well-established chaos detection technique, the SALI method. Our analysis shows that both the $D_L^n$ and the $R_L^n$ indices faced problems in correctly revealing the nature of some orbits, such as sticky chaotic orbits at the borders of stability islands. 
The identification of sticky orbits is of significance in diverse models, like  systems describing chemical reactions~\cite{Naik2020} or the motion of stars in galactic potentials.~\cite{Contopoulos2013,Patsis2022} Consequently, the use of the  DNLD and RNLD indices to detect such orbits, possibly in concert with focused SALI computations, is a topic deserving further investigation.
In addition, we predicted and numerically verified that the DNLD index performs better for systems with large, extended chaotic regions, while the RNLD method exhibits better behavior for systems with larger regular regions. Nevertheless, despite these shortcomings we found that both indices show an overall very good performance, as their classifications are in accordance with the ones obtained by the SALI at a level of at least $90\%$  agreement. 

Studying the effect of the final integration time  (or the total number of map iterations) and the grid spacing of neighboring ICs on the performance of the DNLD and RNLD indices in~Sect.~\ref{sub:effect_of_parameters}, we found that even relatively short (but not too short) integration times of coarse grid LD computations are sufficient to provide a reliable description of the dynamics. Furthermore, taking into account that the evaluation of the DNLD and  RNLD indices (which was based only on the forward in time computation of LDs) typically required one third of the CPU time needed for implementing the SALI method, we realize that the proposed techniques not only provide a clear visualization of phase space structures, but in fact can be used to quantify chaos in a very accurate and efficient manner in continuous and discrete systems.

Our work constitutes a first step in investigating approaches of using information gained by LDs computations for identifying chaotic behavior in dynamical systems, without focusing on the visualization of phase space structures. The development and refinement of methods of this kind is expected to be useful also for high dimensional systems, where the global visualization of their phase space is not possible, due to the space's high dimensionality.
The application of the DNLD and RNLD indices to dynamical systems of higher dimensionality is a natural extension of the current work, and will form part of the future steps of our investigations.

\begin{figure*}
    \centering
    \includegraphics[width=0.99\textwidth]{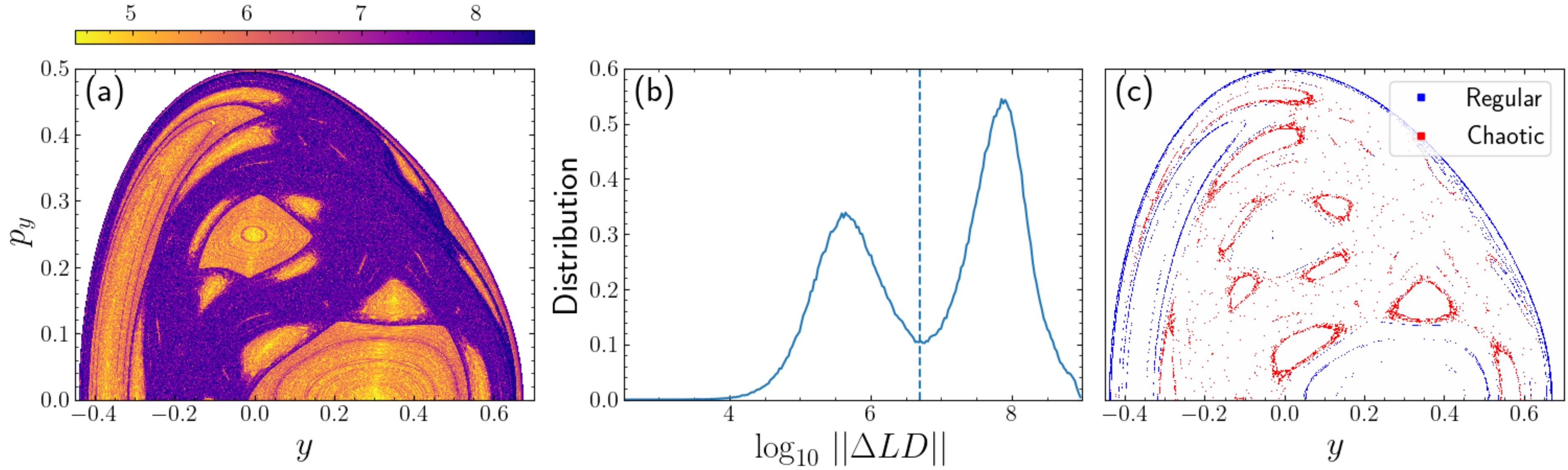}
    \caption{Results obtained for the case of Fig.~\ref{fig:2d_125_cha_reg} using the $||\Delta LD||$ index.  (a) Orbits' ICs colored according  $\log_{10} ||\Delta LD||$ using the color scales at the top of the panel. (b) Normalized distributions of the  $\log_{10}||\Delta LD||$ values of the orbits in (a), with $\log_{10} \alpha_{\Delta}=6.7 $  denoted by a vertical blue dashed line. (c) The set of regular (blue points) and chaotic (red points) ICs which are incorrectly characterized by the $||\Delta LD||$ index.   }
    \label{fig:2dmap_2}
\end{figure*}

Following the directions we set up in this work, we can also implement other quantities to discriminate between chaotic and regular motion, such as the recently presented  $||\Delta LD||$ method,~\cite{Daquin2022} in the same way as we used the DNLD and the RNLD indices. The $||\Delta LD||$ index relies on numerical evaluations of the second spatial derivative of the LDs to visualize the phase space structure. Assuming a similar set up to that used for the definitions of the $D^n_L$ \eqref{eq:defDNLD} and the $R^n_L$ \eqref{eq:defRNLD} indicators, the $||\Delta LD||$ index on a one-dimensional grid of ICs can be estimated by using the second symmetric derivative formula as
\begin{equation}
    \label{eq:DELTALD}
    || \Delta LD || (\mathbf{x})= \displaystyle \frac{\left|M(\mathbf{y}_i^+)-2M(\mathbf{x}) + M(\mathbf{y}_i^-)\right|}{\sigma^2},
\end{equation}
while an analogous approach can be used for higher dimensional subspaces of the system's phase space. The results of Fig.~\ref{fig:2dmap_2} demonstrate that this quantity can be used in the same manner we implemented the DNLD and the RNLD methods, as color maps of its values show the phase space structure of the  H\'enon--Heiles system [Fig.~\ref{fig:2dmap_2}(a)]. Furthermore, using the distribution of the index values [Fig.~\ref{fig:2dmap_2}(b)] to determine a threshold value ($\log_{10} \alpha_{\Delta}=6.7$) for discriminating between regular and chaotic motion, we obtain an accuracy of $P_A \approx 94.2 \%$ with respect to the SALI classification [the ICs of the incorrectly classified orbits are seen in  Fig.~\ref{fig:2dmap_2}(c)]. 

As a final comment let us note that the DNLD and  RNLD methods are not intended to be very precise in estimating the chaotic part of a dynamical system. This task can be performed by well-established and efficient methods which have been designed exactly for that purpose, like the SALI we used in our study. The main advantage of the approaches presented here reside in their ability to provide a reliable estimation of the overall chaoticity of regions in phase space from computationally relatively cheap and simple short time calculations. With large grid spacings and short integration times typically used for LD computations, by applying the DNLD and  RNLD indicators to a preexisting set of LD values, which could have been computed independently for visualizing phase space structures, we can obtain a very useful quantitative byproduct, a trustworthy estimation of the extent of chaos in an ensemble of orbits. 

\begin{acknowledgments}
M.~H.~acknowledges support by the National Research Foundation (NRF) of South Africa (grant number 132726). A.N. acknowledges support from the University of Cape Town (University Research Council, URC) postdoctoral Fellowship grant and support from the Oppenheimer Memorial Trust (OMT). We thank the High Performance Computing facility of the University of Cape Town and the Centre for High Performance Computing \cite{chpc} of South Africa for providing computational resources for this project.
\end{acknowledgments}

\section*{Author Declarations}
The authors have no conflicts to disclose.

\section*{Data Availability Statement}
The data that support the findings of this study are available within the article, and upon reasonable request.


\nocite{*}
\bibliography{HZNKWS}

\end{document}